\documentclass{aa}  
\usepackage{multirow}
\usepackage{graphicx}
\usepackage{threeparttable}
\usepackage{color}
\usepackage{ulem}

\usepackage[colorlinks,citecolor=blue,urlcolor=blue,linkcolor=blue]{hyperref} 
\begin{document} 
\title{Witnessing the fragmentation of a filament into prestellar cores in Orion~B/NGC~2024\thanks{Partly based on observations carried out with the  IRAM NOEMA Interferometer under project number W16AC. IRAM is supported by INSU/CNRS (France), MPG (Germany) and IGN (Spain).}}

   \author{
   Y. Shimajiri\inst{1,2,3}, 
   Ph. Andr$\acute{\rm e}$\inst{3}, 
   N. Peretto\inst{4}, 
   D. Arzoumanian\inst{2,5,6}, 
   E. Ntormousi\inst{7,8}, 
   and
   V. K$\ddot{\rm o}$nyves\inst{9} 
   %and,
   %B. Ladjelate\inst{8}
   }
\institute{  
\inst{1} Kyushu Kyoritsu University, Jiyugaoka 1-8, Yahatanishi-ku,Kitakyushu, Fukuoka, 807-8585, Japan\\
\email{y-shimajiri@fains.jp}\\   
\inst{2} National Astronomical Observatory of Japan, Osawa 2-21-1, Mitaka, Tokyo 181-8588, Japan\\  
\inst{3} Laboratoire d'Astrophysique (AIM), Universit\'e Paris-Saclay, Universit\'e Paris Cit\'e, CEA, CNRS, AIM, 91191 Gif-sur-Yvette, France \\
\inst{4} Cardiff University, School of Physics \& Astronomy, Queen’s buildings, The parade, Cardiff CF24 3AA, UK\\
\inst{5} Instituto de Astrofisica e Ciencias do Espaco, Universidade do Porto, CAUP, Rua das Estrelas, PT4150-762 Porto, Portugal\\
\inst{6} Aix Marseille Univ, CNRS, CNES, LAM, Marseille, France
\inst{7} Scuola Normale Superiore, Piazza dei Cavalieri 7, 56126 Pisa\\
\inst{8} Foundation for Research and Technology (FORTH), Nikolaou Plastira 100, Vassilika Vouton, GR – 711 10 Heraklion, Greece\\
\inst{9} Jeremiah Horrocks Institute, University of Central Lancashire, Preston PR1 2HE, UK\\
}
   \date{Received; accepted}
% \abstract{}{}{}{}{} 
% 5 {} token are mandatory
 
  \abstract
{Recent $Herschel$ observations of nearby molecular clouds have shown that filamentary structures are ubiquitous and that most prestellar cores form in dense filaments. Probing the detailed density and velocity structure of molecular filaments is therefore crucial for improving our observational understanding of the star formation process.
   }
{We aim to characterize both the density and the velocity field of a typical molecular filament in the process of fragmenting into cores.}
{We mapped a portion of the NGC~2024 region in the Orion~B molecular cloud with the Nobeyama 45m telescope, in the $^{12}$CO ($J$=1--0), $^{13}$CO ($J$=1--0), C$^{18}$O ($J$=1--0), and H$^{13}$CO$^+$ ($J$=1--0) lines, and the southwestern part of NGC~2024, corresponding to the NGC~2024S filament, with the NOrthern Extended Millimeter Array (NOEMA) interferometer in H$^{13}$CO$^+$ ($J$=1--0).} 
{The maps of $^{13}$CO, C$^{18}$O, and H$^{13}$CO$^+$ emission trace at least part of the filamentary structure seen in the 8$\arcsec$-resolution ArT$\acute{\rm e}$MiS+{\it Herschel} data. The median radial column density profile of the NGC~2024S filament as derived from ArT$\acute{\rm e}$MiS+$Herschel$ dust emission data is well fitted by a Plummer profile with a half-power diameter $D_{\rm HP}^{\rm Plummer}$=\,0.081$\pm$0.014 pc, which is similar to the findings of previous studies of nearby molecular filaments with $Herschel$. On the other hand, the half-power diameters of NGC~2024S as measured from the Nobeyama $^{13}$CO and C$^{18}$O data are broader, while the half-power diameter derived from the H$^{13}$CO$^+$ data is narrower than the filament diameter measured with $Herschel$. These results suggest that the $^{13}$CO and C$^{18}$O data trace only the (low-density)  outer part of the $Herschel$ filament and the H$^{13}$CO$^+$ data only the (dense) inner part. We identified four cores in the portion of the $Herschel$ map covered by NOEMA and found that each $Herschel$ core corresponds to a single core detected in the combined NOEMA$+$45m H$^{13}$CO$^+$ data cube. The Nobeyama H$^{13}$CO$^+$ centroid velocity map reveals velocity gradients along both the major and the minor axis of the NGC~2024S filament, as well as velocity oscillations with a period $\lambda$ $\sim$0.2 pc along the major axis. Comparison between the centroid velocity and the column density distribution shows a tentative $\lambda$/4 phase shift in H$^{13}$CO$^+$ or C$^{18}$O. This $\lambda$/4 shift is not simultaneously observed for all cores in any single tracer but is tentatively seen for each core in either H$^{13}$CO$^+$ or C$^{18}$O. The difference between the H$^{13}$CO$^+$ and C$^{18}$O velocity patterns may arise from differences in the range of densities probed by H$^{13}$CO$^+$ and C$^{18}$O. We produced a toy model taking into account the three velocity-field components: a transverse velocity gradient, a longitudinal velocity gradient, and a longitudinal oscillation mode caused by fragmentation. Examination of synthetic data shows that the longitudinal oscillation component produces an oscillation pattern in the velocity structure function of the model. Since the velocity structure function of the Nobeyama H$^{13}$CO$^+$ centroid velocity data does show an oscillation pattern, we suggest that our observations are partly tracing core-forming motions and fragmentation of the NGC~2024S filament into cores. We also found that the mean core mass in NGC~2024S corresponds to the effective Bonnor-Ebert mass in the filament. This is consistent with a scenario in which higher-mass cores form in higher line-mass filaments. }
{}

\keywords{
ISM: individual objects:NGC~2024--
ISM: clouds -- stars:formation
               }
\titlerunning{Fragmentation of the NGC~2024S filament into cores}
\authorrunning{Shimajiri et al.}

\maketitle
%
% * <yoshito1024@hotmail.com> 2018-04-16T08:55:35.803Z:
%
% ^.
%________________________________________________________________
%%%%%%%%%%%%%%%%%%%%%
% Section 1
%%%%%%%%%%%%%%%%%%%%%
\section{Introduction}\label{Sect1}

Molecular clouds have long been known to exhibit long filamentary structures \citep[e.g.,][]{Schneider79}. {\it Herschel} observations have confirmed that such filaments are truly ubiquitous in the cold interstellar medium (ISM) of the Milky Way \citep[e.g.,][]{Molinari10, Arzoumanian11, Arzoumanian19, Palmeirim13, Cox16, Schisano20}. Filaments are observed in both actively star-forming and quiescent, non-star-forming molecular clouds \citep[cf.][]{Andre10,Miville-Deschenes10}. The {\it Herschel} observations also show that the typical width of nearby ($< 500\,$pc) filaments measured in H$_2$ column density maps is $\sim$0.1~pc with a factor of $\sim$2 dispersion around this value \citep{Arzoumanian11,Arzoumanian19}.  There has been some debate about the reliability of this finding \citep{Panopoulou+2017, Panopoulou+2022,Hacar+2022}, but tests performed on synthetic data  suggest that {\it Herschel} width measurements are free from significant biases, at least in the case of nearby, high-contrast filamentary structures \citep{Roy+2019,Arzoumanian19,Andre+2022}. While identifying a robust theoretical model for the origin of this typical filament width has been difficult \citep[e.g.][]{Hennebelle13,Ntormousi16}, a promising albeit incomplete idea suggests a connection with the magneto-sonic scale of interstellar turbulence in diffuse molecular gas  \citep[][]{Federrath16}. Given the debate, it is very valuable to keep exploring filament widths further with new observational analyses.

The formation mechanism of molecular filamentary structures is not fully understood \citep[cf.][for a review]{Pineda+2022}, but there is some evidence that molecular filaments may form and grow within sheet-like structures resulting from compression of interstellar matter by large-scale shock waves \citep[][]{Palmeirim13,Arzoumanian18nro,Shimajiri18,Chen20,Bonne20}. Observations with {\it Herschel}  have also shown that most prestellar cores are embedded within dense molecular filaments \citep[e.g.][]{Andre10,Konyves15,Konyves19,Marsh16}, suggesting that molecular filaments are the main sites of at least low- to intermediate-mass star formation. In a particular case in Taurus, direct kinematic evidence of core-forming motions along a filament has even been reported, thanks to observations with the IRAM 30m telescope, in the form of coherent velocity and density oscillations with a $\lambda$/4 phase shift between the density and the velocity field \citep{Hacar11}. Mostly based on the {\it Herschel} results, \citet{Andre14} proposed a scenario for star formation in filaments whereby large-scale compression of interstellar material in supersonic flows generates a complex web of $\sim$0.1-pc-wide filaments in the cold ISM and these filaments then fragment into prestellar cores by gravitational instability. This scenario has the merit that it may possibly account for the ``base'' of the prestellar core mass function (CMF) and by extension the stellar initial mass function (IMF) for $0.1\, M_\odot \la M_\star \la 1\, M_\odot  $. In particular, there is evidence that molecular filaments may fragment in qualitatively the same manner at low and high masses \citep[cf.][]{Shimajiri19b} and that the prestellar CMF may be partly inherited from the distribution of filament line masses \citep{Andre19}. The validity and details of this filament scenario for star formation and the IMF are actively debated, however \citep[see, e.g.,][]{Gong15}. But beyond on-going debates, there is little doubt after the {\it Herschel} results that dense filaments are representative of the initial conditions of the bulk of star formation in molecular  clouds. Characterizing the fragmentation mechanism of molecular filaments and their detailed density and velocity structure is thus crucial to our understanding of the star formation process. 

Here, in an effort to clarify how prestellar cores form and grow within filaments, we present a detailed fragmentation study of the intermediate-mass filament NGC~2024S in Orion~B, which has a line mass $M_{\rm line}$ of $\sim$62$\pm$13 $M_{\odot} {\rm pc}^{-1}$\footnote{The line mass was estimated by integrating over the filament area corresponding to $A_{\rm V}$ > 8 in the {\it Herschel} column density map where the back-ground emission is  subtracted. The uncertainty comes from the uncertainty in the background subtraction.} The line mass $M_{\rm line}$ of $\sim$62 $M_{\odot} {\rm pc}^{-1}$ exceeds the thermally critical line mass $\sim$16 $M_{\odot}\, {\rm pc}^{-1}$ of an isothermal filament at $\sim$10\,K, suggesting that the NGC~2024S filament may not be in radial equilibrium. If this is the case, radial perturbations are expected to grow faster than perturbations along the filament axis, implying that the filament may not be able to fragment into prestellar cores before radially contracting to a spindle \citep[e.g.,][]{Inutsuka92,Inutsuka97}. However,  magneto hydrodynamics (MHD) turbulence and/or static magnetic fields can increase the effective critical line mass \citep{Fiege00, Jackson10,Tomisaka14,Kashiwagi21,Pattle22}. Accordingly, radial support provided by MHD waves and/or magnetic fields may stabilize the filament and allow it to fragment along its length. 

The NGC~2024 region is located in the southern part of the Orion~B molecular cloud \citep[$d$ = 400 pc, ][]{Gibb08} and is known to be a very active site of star formation, with an estimated star formation rate of 9.2--13.8 $\times$10$^{-6}$ $M_{\odot}$ yr$^{-1}$ \citep{Shimajiri17}. Using {\it Herschel}  Gould Belt Survey (HGBS) data, \citet{Konyves19} found that 60-90\% of prestellar cores are closely associated with filaments in Orion~B and observed that the most massive prestellar cores are spatially segregated in the highest column density areas. Orion~B including the NGC~2024 region was also observed as part of the  ORION-B (Outstanding Radio Imaging of OrioN~B) large program with the IRAM~30m telescope in 15 molecular lines such as $^{12}$CO (1--0), $^{13}$CO (1--0), C$^{18}$O (1--0), and H$^{13}$CO$^+$(1--0) with a velocity resolution of $\sim$0.5 km s$^{-1}$ \citep{Pety16}. These authors reported that 54.5\%, 39.4\%, 23.5\%, 7.8\% of the total line fluxes in $^{12}$CO (1--0), $^{13}$CO (1--0), C$^{18}$O (1--0), and H$^{13}$CO$^+$ (1--0) are from the $A_{\rm V}$=1--6 area, while 45.6\%, 60\%, 78\%, 90\% of the same line fluxes are from the $A_{\rm V}$=6--222 area in Orion~B, respectively. Based on the ORION-B C$^{18}$O (1--0) data, \citet{Orkisz19} found an average filament width of  $\sim$0.12$\pm$0.04 pc,  consistent with the typical filament width found from {\it Herschel} column density data \citep[e.g.,][]{Arzoumanian19}. Recently, the presence of a cloud-cloud collision in this region was suggested by NANTEN2 $^{13}$CO(2--1) observations \citep{Enokiya19}.

To investigate how molecular filaments fragment into cores, we performed observations of NGC~2024 with both the Nobeyama 45m telescope and the NOrthern Extended Millimeter Array (NOEMA) interferometer. In this paper, we focus on the southwestern part of NGC 2024, NGC~2024S, to avoid the effect of the H$_{\rm II}$ region located in the northern part of NGC 2024. The paper is organized as follows: in Sect. \ref{Sect2}, we describe our Nobeyama 45m and NOEMA observations. In Sect. \ref{Sect3}, we present the results of $^{12}$CO (1--0), $^{13}$CO (1--0), C$^{18}$O (1--0), and H$^{13}$CO$^+$ (1--0) mappings toward the Orion~B/NGC~2024 region. In Sect. \ref{Sect4}, we discuss whether the cores are formed via the fragmentation of the filament. In Sect. \ref{Sect5}, we summarize our results.

%%%%%%%%%%%%%%%%%%%%%%%%%%%%%%%%%%%%%%%%%%
% Section 2 : Observations
%%%%%%%%%%%%%%%%%%%%%%%%%%%%%%%%%%%%%%%%%%
\section{Observations and data}\label{Sect2}

\subsection{Herschel GBS and ArT$\acute{\rm e}$MiS+Herschel column density maps}

\begin{figure*}
\centering
\includegraphics[angle=0,width=18cm]{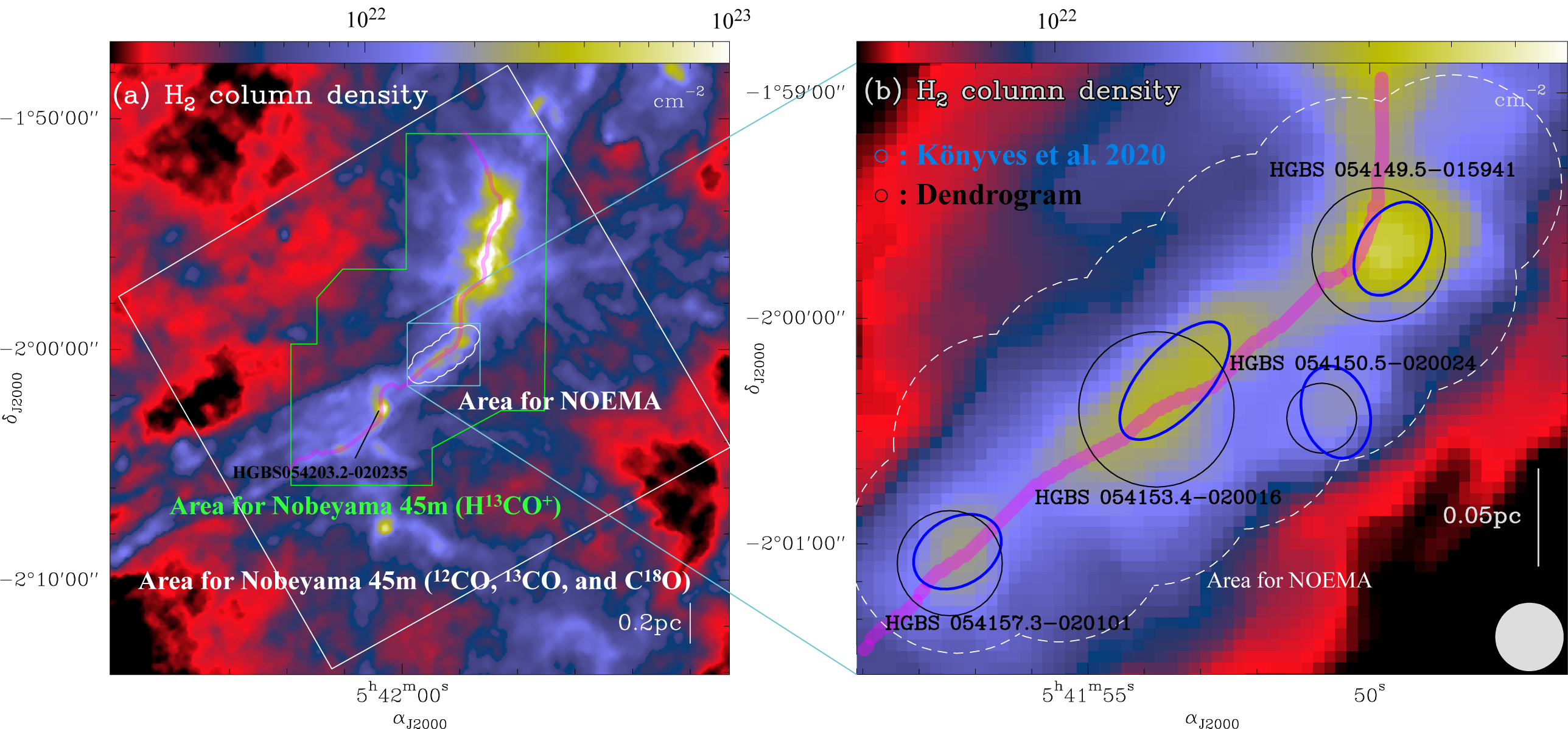}
\caption{$Herschel$ column density map at an angular resolution of 18$\arcsec$.2  toward the NGC~2024 region ($a$) and toward the area observed with the NOEMA interferometer ($b$). The $Herschel$ column density map is from HGBS data (\citealp{Konyves19}). In both panels, the magenta curve indicates the filament crest. The filament crest is determined by DisPerSE algorithm \citep{Sousbie11,Sousbie11b, Arzoumanian11}. In panel ($a$), the white box indicates the field observed in the $^{12}$CO (1--0), $^{13}$CO (1--0), and C$^{18}$O (1--0) lines with the Nobeyama 45m telescope. The green polygon outlines the field observed in H$^{13}$CO$^+$ (1--0). The dashed white circles indicate the field of view of the NOEMA mosaic observations. In panel ($b$), the blue open ellipses mark the cores identified by \citet{Konyves19} and the black open circles the cores identified in the $Herschel$ map by the dendrogram analysis. The sizes of the ellipses and circles reflect the core sizes estimated by  \citet{Konyves19} and by the dendrogram analysis, respectively.
}
\label{fig1}
\end{figure*}

We used the {\it Herschel} H$_2$ column density map constructed from HGBS data by \citet{Konyves19}, publicly available\footnote{\url{http://gouldbelt-herschel.cea.fr/archives}}. The effective resolution of this column density map is 18$\arcsec $.2. Figure \ref{fig1} shows the {\it Herschel} column density map toward the Orion~B/NGC~2024 region. 

We also produced a higher resolution column density map by combining the {\it Herschel} data with  ArT$\acute{\rm e}$MiS data following the approach described in \citet[][]{Schuller21}. Hereafter, we call this map the ArT$\acute{\rm e}$MiS+Herschel column density map. The details of the Orion~B/NGC~2024 ArT$\acute{\rm e}$MiS observations, similar to the Orion~A observations presented by \citet[][]{Schuller21} will be given in a separate paper.

%%%%%%%%%%%%%%%%%%%%%%%%%%%%%%%%%%%%%%%%%%
% Section 2.1 : Nobeyama observations
%%%%%%%%%%%%%%%%%%%%%%%%%%%%%%%%%%%%%%%%%%

\subsection{Nobeyama 45m observations}

%%%%%%%%%%%%%%%%%%%%%%%%%%%%%%%%%%%%%%%%%%%%%%%%%%%%%%%%%%%%%%%
% Section 2.1.1 : Nobeyama 12CO, 13CO, C18O observations
%%%%%%%%%%%%%%%%%%%%%%%%%%%%%%%%%%%%%%%%%%%%%%%%%%%%%%%%%%%%%%%

Between 27 February 2017 and 1 March 2017, we carried out mapping observations in $^{12}$CO (1--0, 115.2701204 GHz), $^{13}$CO (1--0, 110.201354 GHz), C$^{18}$O (1--0, 109.782176 GHz), and H$^{13}$CO$^+$ (86.75433 GHz) toward the NGC~2024 region in the Orion~B molecular cloud with the FOREST receiver installed on the Nobeyama 45m telescope. The $^{12}$CO (1--0), $^{13}$CO (1--0), and C$^{18}$O (1--0) lines were observed simultaneously. At 115 GHz, the telescope has a beam size of 15$\arcsec$.1 (HPBW). As the backend, we used the spectrometer, Spectral Analysis Machine for the 45m telescope (SAM45), which has a 31 MHz bandwidth and a frequency resolution of 7.63 kHz. The frequency resolution corresponds to a velocity resolution of $\sim$0.02 km s$^{-1}$ at 115 GHz. The standard chopper wheel method was used to convert the observed signal to the antenna temperature $T_{\rm A}^*$ in units of K, corrected for the atmospheric attenuation. To calibrate the intensity scale for the CO (and isotopes), we observed FIR 4 in OMC-2 with a small box of  2$\arcmin \times 2 \arcmin$ and a center of (RA$_{\rm J2000}$, DEC$_{\rm J2000}$)=(5$^{\rm h}$35$^{\rm m}$26$^{\rm s}$.8, $-5^{\circ}$9$^{\rm m}$57$^{\rm s}$.4). By direct comparison between the obtained data and the data obtained in \citet{Shimajiri11, Shimajiri14, Shimajiri17}, we obtained the intensity scaling factors from $T_{\rm A}^*$ to $T_{\rm MB}$ for each line. The estimated intensity scaling factors are applied to all of the data. Thus, the intensities of the data in this paper are in $T_{\rm MB}$. The telescope pointing was checked every hour by observing the SiO maser sources, Ori-KL, and was better than 3$\arcsec$ throughout the entire observation. Our mapping observations were made with the on-the-Fly (OTF) mapping technique. We chose the position of (RA$_{\rm J2000}$, DEC$_{\rm J2000}$) = (5$^{\rm h}$49$^{\rm m}$45$^{\rm s}$.115, $-1^{\circ}$56$^{\rm m}$8$^{\rm s}$.54) as the off position. We obtained OTF maps with two different scanning directions along the RA or Dec axes covering the 20$\arcmin$$\times$20$\arcmin$ for CO and its isotopes and 5$\arcmin$$\times$6$\arcmin$ for H$^{13}$CO$^+$ and combined them into a single map to reduce the scanning effects as much as possible. As a convolution function, we applied a spheroidal function with an FWHM of half of the beam size, resulting in an effective beam size of 21$\arcsec$.6 for CO and 25$\arcsec$ for H$^{13}$CO$^+$. In order to improve the sensitivity, we combined the H$^{13}$CO$^+$ data with the data obtained in \citet{Shimajiri17}. We smoothed the data with a Gaussian function resulting in a final effective beam size of 25$\arcsec$ for CO and its isotope and 30$\arcsec$ for H$^{13}$CO$^+$. The 1$\sigma$ noise level of the final data with an effective resolution of 25$\arcsec$ is 0.57 K, 0.30 K, 0.30 K in $T_{\rm MB}$ for $^{12}$CO (1--0), $^{13}$CO (1--0), and C$^{18}$O (1--0) at a velocity resolution of 0.1 km s$^{-1}$ (Table \ref{table1}). The 1$\sigma$ noise level with an effective resolution of 30$\arcsec$ is 0.13 K in $T_{\rm MB}$ for H$^{13}$CO$^+$ (1--0) at a velocity resolution of 0.1 km s$^{-1}$ (Table \ref{table1}).

%Figure 1

%Table 1
%\input{table1.tex}
\begin{table*}
\centering
\begin{threeparttable}
\caption{Nobeyama 45m observations\label{table1}}
\begin{tabular}{|l|cccc|}
\hline
Molecule         &  $^{12}$CO    & $^{13}$CO     & C$^{18}$O     & H$^{13}$CO$^{+}$  \\
Transition       & (1--0)        & (1--0)        & (1--0)        & (1--0)\\
Frequency   & 115.2701204 GHz   & 110.201354 GHz    & 109.782176 GHz    & 86.75433 GHz \\
\hline
Telescope        &\multicolumn{4}{c|}{Nobeyama 45m} \\
Receiver         & FOREST        & FOREST        & FOREST        & TZ\tnote{$^\dag$}  \\
                                 & \multicolumn{3}{c}{}          &  FOREST \\
Spectrometer     & \multicolumn{4}{c|}{SAM45} \\
Obs. period      & \multicolumn{3}{c}{28 Feb. -- 1 March 2017}   &   7-21 May 2015\tnote{$^\dag$} \\
                                 & \multicolumn{3}{c}{}          & 27 Feb. 2017  \\
$\theta_{\rm eff}$  
                 &  25$\arcsec$  & 25$\arcsec$   & 25$\arcsec$   & 30$\arcsec$ \\
                 & $\sim$0.05 pc & $\sim$0.05 pc & $\sim$0.05 pc & $\sim$0.06 pc \\
dV        & 0.1 km s$^{-1}$           & 0.1 km s$^{-1}$           & 0.1 km s$^{-1}$           & 0.1 km s$^{-1}$ \\
rms          & 0.57 K          & 0.30 K          & 0.30 K          & 0.13 K \\
\hline
\end{tabular}
\begin{tablenotes}
\item[$^\dag$] The data was taken from \citet{Shimajiri17}.
\end{tablenotes}
\end{threeparttable}
%}
\end{table*}

%Figure 2
\begin{figure*}
\centering
\includegraphics[angle=0,width=18.5cm]{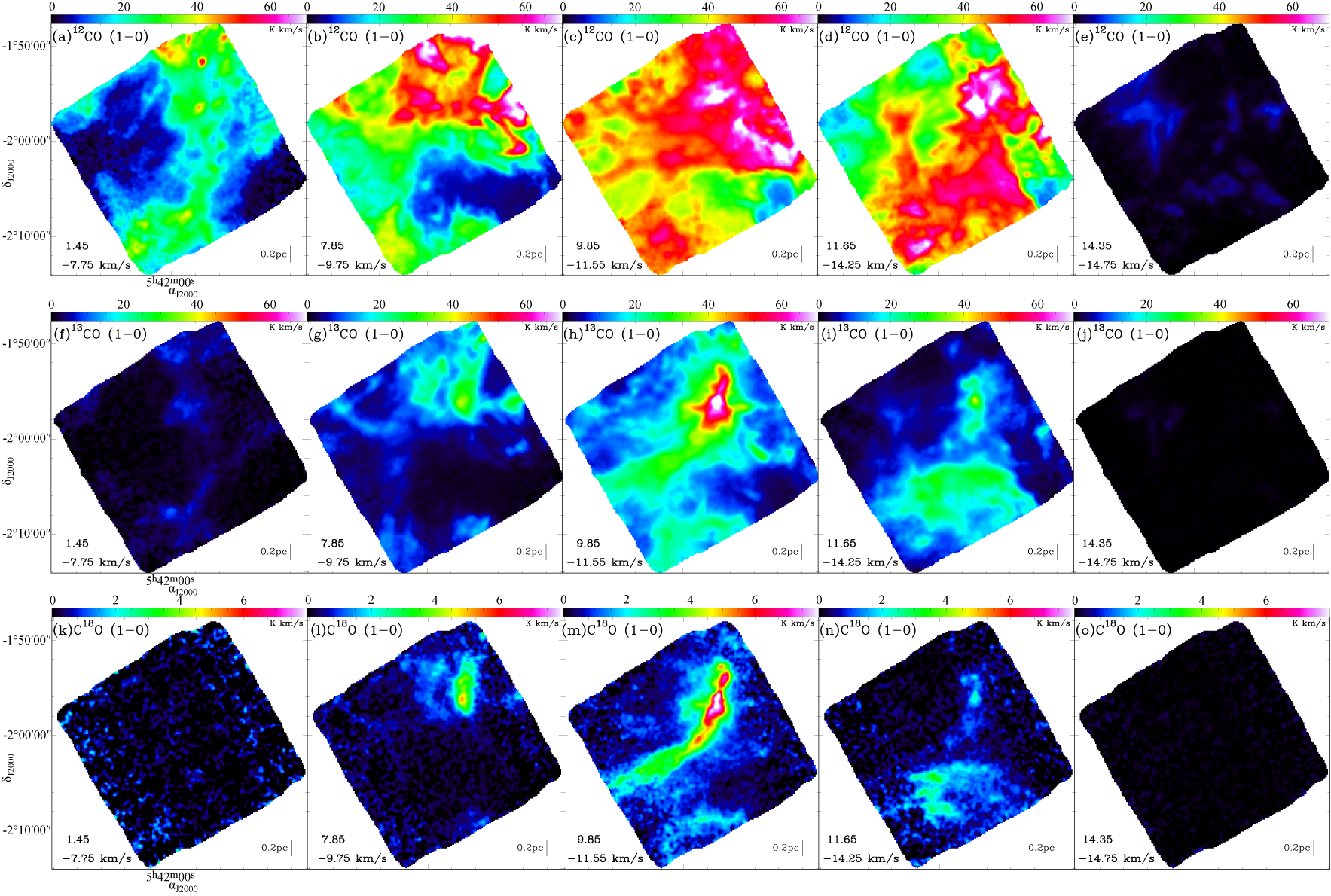}
\caption{($a-e$) $^{12}$CO (1--0), ($f-j$) $^{13}$CO (1--0), and ($k-o$) C$^{18}$O (1--0) maps integrated in the velocity ranges of 1.45-7.75 km s$^{-1}$, 7.85--9.75 km s$^{-1}$, 9.85-11.55 km s$^{-1}$, 11.65-14.25 km s$^{-1}$, and 14.35--14.75 km s$^{-1}$. The integrated velocity range is indicated in the bottom-left corner of each $panel$. The coverage of these CO and its isotope observations is also shown in Fig. \ref{fig1}.}
\label{fig2}
\end{figure*}

\begin{figure}
\centering
\includegraphics[angle=0,width=8.5cm]{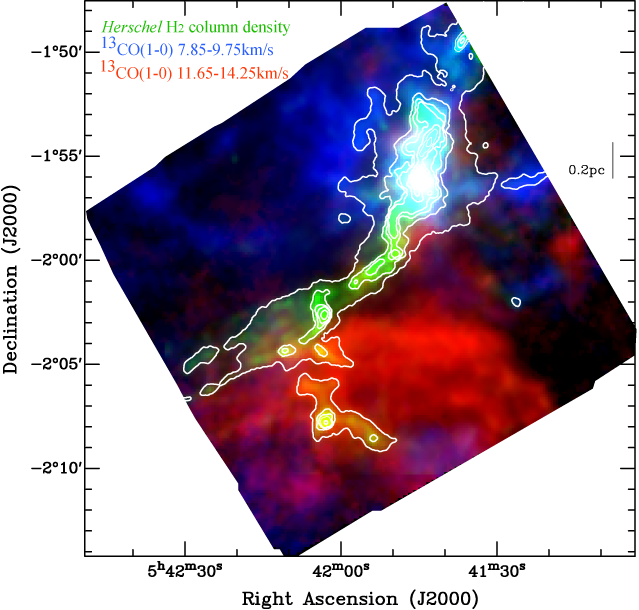}
\caption{A three-color composite image of the {\it Herschel} H$_2$ column density map (green) and the blue- and red-shifted Nobeyama $^{13}$CO emission (red: 7.85--9.75 km s$^{-1}$ and blue:11.65--14.25 km s$^{-1}$, see also Figs. \ref{fig2}g and \ref{fig2}i). The white contours correspond to $A_{\rm V}$ levels of 8, 16, 24, 32, 64, 128, and 256 mag (assuming $N_{\rm H_2}$ /$A_{\rm V}$ = 0.94 $\times$ 10$^{21}$ cm$^{-2}$, \cite{Bohlin78}) in the {\it Herschel} H$_2$ column density map at an angular resolution of 18$\arcsec$.2.
}
\label{fig3}
\end{figure}

%%%%%%%%%%%%%%%%%%%%%%%%%%%%%%%%%%%%%%%%%%
% Section 2.2 : NOEMA observations
%%%%%%%%%%%%%%%%%%%%%%%%%%%%%%%%%%%%%%%%%%
\subsection{NOEMA observations}

We carried out millimeter interferometric 12-pointing mosaic observations of the NGC~2024 region in the H$^{13}$CO$^+$ (1--0) line at 86.75433 GHz with the NOrthern Extended Millimeter Array (NOEMA) in the D configuration during a period from 9 August 2016 to 1 September 2017. The data were obtained with the narrow-band correlator which was configured with 512 channels per baseline and a bandwidth of 20 MHz.  The channel spacing is 39 kHz which corresponds to 0.13 km s$^{-1}$ at 86 GHz. Table \ref{table2} summarizes the parameters for the line observations. Using CLIC which is part of the GILDAS software, calibration was carried out following standard procedures. We adopted natural weighting for the imaging of the H$^{13}$CO$^+$ emission. Since the minimum projected baseline length of the H$^{13}$CO$^+$ observations was 4.5 k$\lambda$, the NOEMA data are insensitive to structures more extended than 36$\arcsec$.7 (0.07 pc) at the 10\% level \citep{Wilner94}. 

%Table 2
%\input{table2.tex}
\begin{table}
\centering
\begin{threeparttable}
\caption{NOEMA observations\label{table2}}
\begin{tabular}{|l|c|}
\hline
Configuration                       & D \\
Baseline                & 4.5 -- 37.0 k$\lambda$\\
Primary beam HPBW        & 58$\arcsec$.1$\sim$0.11 pc\\
Synthesized Beam HPBW    & 6$\arcsec$.40 $\times$ 3$\arcsec$.68 (PA: -186.18$^{\circ}$) \\
                                    & 0.012 pc $\times$ 0.007 pc \\
Velocity resolution    & 0.13 km s$^{-1}$ \\
Gain calibrators                    & 0458-020, 0550+032 \\
Bandpass calibrator                 & 3C84\\
Rms noise level  & 16 mJy beam$^{-1}$\\
\hline
\end{tabular}
\end{threeparttable}
\end{table}

%Figure 4
\begin{figure*}
\centering
\includegraphics[angle=0,width=14cm]{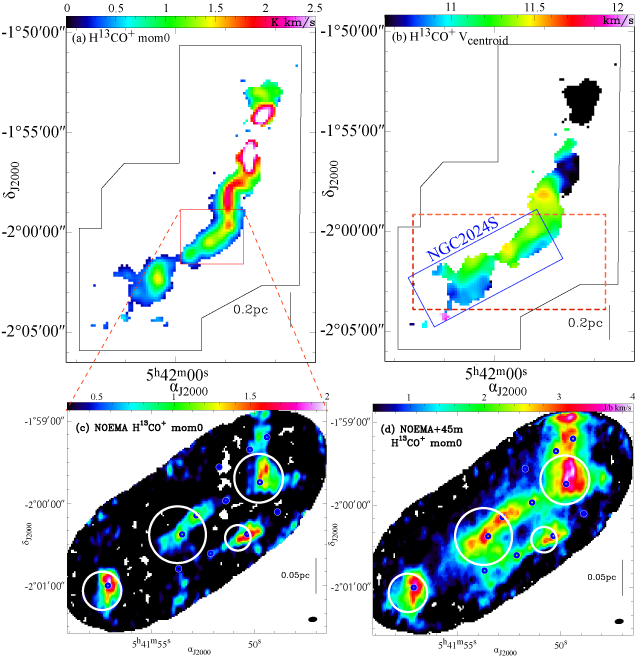}
\caption{($a$) Nobeyama 45m H$^{13}$CO$^+$ (1--0) integrated intensity, ($b$) H$^{13}$CO$^+$ (1--0) centroid velocity map, ($c)$ NOEMA, and ($d$) NOEMA+45m H$^{13}$CO$^+$ (1--0) integrated intensity maps. The integrated velocity range is from 10.21 km s$^{-1}$ to 11.96 km s$^{-1}$. In $panels$ a and b, black polygons outline the field observed in the H$^{13}$CO$^+$ (1--0) line. In $panel$ a, the red box indicates the area shown in $panels$ c and d. In $panel$ b, the red dashed box indicates the area shown in Fig. \ref{fig7}. The white open circles in the panels c and d indicate the positions of the cores identified in the $Herschel$ map by the dendrogram analysis. The sizes of the white open circles reflect the {\it Herschel} source sizes. The small blue open circles in $panels$ c and d mark the positions of the cores identified in the NOEMA H$^{13}$CO$^+$ map by the dendrogram analysis with a fixed symbol size. The filled circles at the bottom right indicate the beam sizes in the panels a and b. See also Fig. \ref{figA1}.
}
\label{fig4}
\end{figure*}

\subsection{Combining NOEMA and 45m H$^{13}$CO$^+$ data}

In order to produce a data set with both information on extended emission and high angular resolution, we regridded the Nobeyama H$^{13}$CO$^+$(1--0) data to NOEMA H$^{13}$CO$^+$(1--0) data both in velocity and position using the task ``regrid'' and combined the NOEMA and Nobeyama H$^{13}$CO$^+$(1--0) observations using the task ``immerge'' in the miriad software package \citep{Sault95}. A calibration factor of 1.0 was applied to the NOEMA H$^{13}$CO$^+$(1--0) data.

Figure \ref{figA2} in Appendix~A compares the velocity channel maps of i) the NOEMA H$^{13}$CO$^+$(1--0) data, ii) the Nobeyama H$^{13}$CO$^+$(1--0) data, iii) the combined NOEMA $+$ Nobeyama data (hereafter, called NOEMA$+$45m data), and iv) the NOEMA$+$45m data smoothed to the angular resolution of the Nobeyama H$^{13}$CO$^+$(1--0) data (hereafter, called the smoothed NOEMA$+$45m data). It can be seen in each channel map that extended emission has been restored in the NOEMA$+$45m data compared to the NOEMA-only data. In addition, the smoothed NOEMA$+$45m data cube is very similar to the Nobeyama data cube: The intensity in the smoothed NOEMA$+$45m data is consistent to within 10\% with that in the Nobeyama data. The rms noise level of the NOEMA$+$45m data at a velocity resolution of 0.13 km s$^{-1}$ is 0.017 Jy beam$^{-1}$.

%%%%%%%%%%%%%%%%%%%%%
% Section 3 :Results
%%%%%%%%%%%%%%%%%%%%%
\section{Results and Analysis}\label{Sect3}

%%%%%%%%%%%%%%%%%%%%%%%%%%%%%%%%%%%%%%%%%%
% Section 3.1 : Results of 12CO, 13CO, C18O
%%%%%%%%%%%%%%%%%%%%%%%%%%%%%%%%%%%%%%%%%%
\subsection{Spatial distribution of $^{12}$CO (1--0), $^{13}$CO (1--0), C$^{18}$O (1--0), and H$^{13}$CO$^+$(1--0) emission}

\subsubsection{$^{12}$CO (1--0), $^{13}$CO (1--0), and C$^{18}$O (1--0) emission}\label{sec:results-CO}

Figure \ref{fig2} shows the velocity integrated intensity maps observed in the $^{12}$CO (1--0), $^{13}$CO (1--0), and C$^{18}$O (1--0) lines at the Nobeyama 45m telescope. In the maps for 1.5 < $V_{\rm LSR}$ < 7.8 km s$^{-1}$ (Figs. \ref{fig2}a, f, and k), faint $^{12}$CO (1--0) and $^{13}$CO (1--0) emission can be seen. In the maps for 7.9 < $V_{\rm LSR}$ < 9.8 km s$^{-1}$  (Figs. \ref{fig2}b, g, and l), strong $^{13}$CO (1--0) and C$^{18}$O (1--0) emission is associated with the main peak in the {\it Herschel} column density map. Furthermore, the $^{12}$CO (1--0) and $^{13}$CO (1--0) emission is stronger in the northern part of the field. In the maps for 9.9 < $V_{\rm LSR}$ < 11.6 km s$^{-1}$ (Figs. \ref{fig2}c, h, and m), the $^{13}$CO (1--0) and C$^{18}$O (1--0) emission is seen toward the filament traced in the {\it Herschel} column density map of Fig. \ref{fig1}. In the maps for 11.7 < $V_{\rm LSR}$ < 14.3 km s$^{-1}$ (Figs. \ref{fig2}d, i, and n), the emission detected in $^{12}$CO (1--0), $^{13}$CO (1--0), and C$^{18}$O (1--0) is distributed mostly in the southern part of the field. In the maps for 14.4 < $V_{\rm LSR}$ < 14.8 km s$^{-1}$ (Figs. \ref{fig2}e, j, and o), significant emission can be seen only in $^{12}$CO. Figure~\ref{fig3} shows a three-color composite image with the {\it Herschel} H$_2$ column density map (in green) and the blue- and red-shifted $^{13}$CO emission detected by the Nobeyama telescope. The blue-shifted emission lies on the northeast of the filament, while the red-shifted emission lies on the southwest of it.

%%%%%%%%%%%%%%%%%%%%%%%%%%%%%%%%%%%%%%%%%%
% Section 3.2 : Results of H13CO+
%%%%%%%%%%%%%%%%%%%%%%%%%%%%%%%%%%%%%%%%%%
\subsubsection{H$^{13}$CO$^+$(1--0) emission}\label{sec:results-centroid}
 
\citet{Shimajiri17} found that the spatial distribution of H$^{13}$CO$^+$(1--0) emission in NGC~2024 is similar to that seen in dust emission in the {\it Herschel} column density maps of the Ophiuchus, Aquila, and Orion~B clouds and that the optical depth of the H$^{13}$CO$^+$(1--0) line in these clouds is low (see Table A.1 in \citet{Shimajiri17}). This suggests that the H$^{13}$CO$^+$(1--0) line is a good tracer of the dense filaments detected with {\it Herschel} and is suitable to investigate their underlying velocity field.  

Figure \ref{fig4} shows the distribution of the H$^{13}$CO$^+$ (1--0) emission. The H$^{13}$CO$^+$ emission traces the dense part of the filament seen in the {\it Herschel} H$_2$ column density well (Figs. \ref{fig1} and \ref{fig4}a), while the C$^{18}$O emission traces larger-scale structures in the {\it Herschel} H$_2$ column density map (Figs. \ref{fig1} and \ref{fig2}m, see also Fig. \ref{figA3}). At the core scale, the H$^{13}$CO$^+$ emission traces well the cores detected in the {\it Herschel} H$_2$ column density map (see Sect. \ref{sect.core_ID}), while the C$^{18}$O emission does not trace some of the {\it Herschel} cores. This is likely due to the depletion of CO molecules onto grains at high density \citep[e.g.,][]{Tafalla04}. A similar result that H$^{13}$CO$^+$ emission traces dense dusty cores better than C$^{18}$O emission was also reported in the Orion~A molecular cloud \citep[$d$ = 400 pc, ][]{Shimajiri15a}. The correlation between H$^{13}$CO$^+$ and {\it Herschel} H$_2$ data has a smaller scatter than that between C$^{18}$O and {\it Herschel} H$_2$, confirming that the H$^{13}$CO$^+$ emission traces well the dense structures seen in the {\it Herschel} H$_2$ column density map (Fig. \ref{figA4}). Thus, the H$^{13}$CO$^+$ emission provides a better probe of the velocity  and density structure of the cores and filaments seen in the {\it Herschel} H$_2$ column density map than C$^{18}$O.

%%%%%%%%%%%%%%%%%%%%%%%%%%%%%%%%%%%%%%%%%%
% Section 3.2.2 : Distribution in high spatial-resolution map
%%%%%%%%%%%%%%%%%%%%%%%%%%%%%%%%%%%%%%%%%%
\subsubsection{Gas distribution in the NOEMA high spatial-resolution maps}\label{sect:high-reso_map}

\begin{figure*}
\centering
\includegraphics[angle=0,width=18cm]{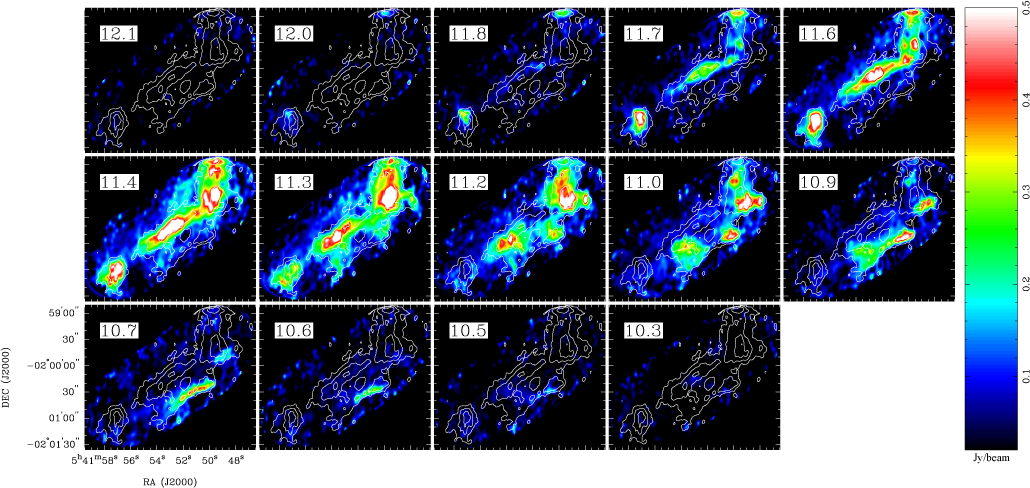}
\caption{Velocity channel maps of NOEMA+45m H$^{13}$CO$^+$ (1--0) emission. Contours indicate NOEMA+45m H$^{13}$CO$^+$ (1--0) integrated intensity map and the contour levels of these maps start at 0.1 Jy beam$^{-1}$ levels with an interval of 0.1 Jy beam$^{-1}$. }
\label{fig5}
\end{figure*}

Panel c of Fig. \ref{fig4} shows the integrated intensity map of the H$^{13}$CO$^+$ (1--0) emission observed with NOEMA. The overall distribution of the NOEMA H$^{13}$CO$^+$ (1--0) emission is consistent with that seen in both the Nobayama H$^{13}$CO$^+$ (1--0) map and the $Herschel$ column density map. In the western part of the NOEMA map (RA$_{\rm J2000}$, DEC$_{\rm J2000}$=$\sim$5$^{\rm h}$41$^{\rm m}$50$^{\rm s}$, $\sim$-2$^{\circ}$0$^{\rm m}$25$^{\rm s}$), a secondary structure can be seen. This structure can also be recognized in the $Herschel$ column density map and in the Nobeyama H$^{13}$CO$^+$(1--0) velocity channel maps at 10.7 < $V_{\rm LSR}$ < 11.0 km s$^{-1}$ (Fig. \ref{fig5}).

%Figure 5

%%%%%%%%%%%%%%%%%%%%%%%%%%%%%%%%%%%%%%%%%%
% Section: Core identification
%%%%%%%%%%%%%%%%%%%%%%%%%%%%%%%%%%%%%%%%%%
\subsubsection{Core identification in the NOEMA$+$45m H$^{13}$CO$^+$ data cube}\label{sect.core_ID}

\citet{Konyves19} obtained a census of dense cores in Orion~B from the $Herschel$ data using the \textsl{getsources} algorithm \citep{Menshchikov12}. In the field mapped here with NOEMA, four cores were identified.

Here, we identified cores in the NOEMA+45m H$^{13}$CO$^+$ data cube to compare core spacing and filament width. As \textsl{getsources} cannot be used with spectral line data,  we performed a dendrogram analysis using the {\it astrodendro} package\footnote{http://www.dendrograms.org} \citep{Rosolowsky08}. When a data set is sensitive to a whole hierarchy of structures such as clumps, filaments, and cores, the dendrogram algorithm is a powerful technique to trace this hierarchy \citep[cf.][]{Friesen16}. In addition, we also extracted cores in the $Herschel$ column density map using the same dendrogram technique for comparison with both the cores identified here in the NOEMA+45m H$^{13}$CO$^+$ data cube and the cores identified by \citet{Konyves19} with \textsl{getsources} in the {\it Herschel} data.

To perform a dendrogram analysis, three input parameters are required. The first one, {\it min$\_$value}, is the starting level, i.e., the minimum intensity value below each extracted structure. The second one, {\it min$\_$delta}, is a step and corresponds to the minimum height of each extracted structure above the starting level. The third one, {\it min$\_$npix}, is the minimum number of pixels that a significant structure must contain. The detected structures are categorized into three types, $trunk$, $branch$, and $leaf$, following their hierarchy \citep[see][]{Rosolowsky08}. 
In this paper, we refer to the smallest,  $leaf$ structures as candidate cores and focus on the detection of such cores. 
 
%Table 6
%\input{table3.tex}
\begin{table*} %Dengrogram_for_12Jan2021.ipynb
\tiny
\centering
\begin{threeparttable}
\caption{$Herschel$ cores identified by {\it Dendrogram} \label{table_core}}
\begin{tabular}{lcccc}
\hline
Core name & HGBS 054157.3-020101 & HGBS 054153.4-020016 & HGBS 054150.5-020024 & HGBS 054149.5-015941 \\
Core Type$^+$ & Prestellar & Prestellar & Prestellar & Class-0/I$^{++}$ \\
\hline
RA$_{\rm J2000}^{\rm Herschel}$ $^{**}$                            & 05$^{\rm h}$41$^{\rm m}$57$^{\rm s}$.3   & 05$^{\rm h}$41$^{\rm m}$53$^{\rm s}$.5  & 05$^{\rm h}$41$^{\rm m}$50$^{\rm s}$.6  & 05$^{\rm h}$41$^{\rm m}$49$^{\rm s}$.6 \\
DEC$_{\rm J2000}^{\rm Herschel}$ $^{**}$                           & -02$^{\circ}$01$^{\rm m}$01$^{\rm s}$.9 & -02$^{\circ}$00$^{\rm m}$16$^{\rm s}$.4 & -02$^{\circ}$00$^{\rm m}$24$^{\rm s}$.7 & -01$^{\circ}$59$^{\rm m}$41$^{\rm s}$.2 \\
$R_{\rm core}^{\rm Herschel}$ [pc]                              & 0.024         & 0.034        & 0.025        & 0.026 \\
$M_{\rm core}^{\rm Herschel}$ $^+$ [$M_{\rm \odot}$]               & 1.63          & 3.80         & 0.99         & 3.59 \\
$dV_{\rm H^{13}CO^{+},FWHM}^{\rm Herschel}$ $^\dag$ [km s$^{-1}$]  & 0.51          & 0.56         & 0.73         & 0.61 \\
$V_{\rm H^{13}CO^{+},sys}^{\rm Herschel}$ $^\dag$ [km s$^{-1}$]    & 11.44         & 11.31        & 10.80        & 11.36 \\
$M_{\rm VIR}^{\rm Herschel}$ $^\ddag$ $[M_{\rm \odot}]$            & 0.79          & 1.35         & 1.68         & 1.19  \\
$\alpha_{\rm VIR}^{\rm Herschel}$ $^*$                             & 0.5           & 0.4          & 1.7          & 0.3 \\
\hline
RA$_{\rm J2000}^{\rm Dendro}$ $^{**}$                           & 05$^{\rm h}$41$^{\rm m}$57$^{\rm s}$.4  & 05$^{\rm h}$41$^{\rm m}$53$^{\rm s}$.7  & 05$^{\rm h}$41$^{\rm m}$50$^{\rm s}$.8  & 05$^{\rm h}$41$^{\rm m}$49$^{\rm s}$.8 \\
DEC$_{\rm J2000}^{\rm Dendro}$ $^{**}$                          & -02$^{\circ}$01$^{\rm m}$03$^{\rm s}$.8 & -02$^{\circ}$00$^{\rm m}$22$^{\rm s}$.9 & -02$^{\circ}$00$^{\rm m}$25$^{\rm s}$.2 & -01$^{\circ}$59$^{\rm m}$41$^{\rm s}$.7 \\
$R_{\rm core}^{\rm Dendro}$ [pc]                             & 0.027        & 0.040        & 0.018        &  0.035 \\
$M_{\rm core}^{\rm Dendro}$ [$M_{\rm \odot}$]                & 1.07         & 2.84         & 0.44         & 2.50 \\
$dV_{\rm H^{13}CO^{+},FWHM}^{\rm Dendro}$ $^\dag$ [km s$^{-1}$] & 0.44         & 0.60         & 0.68         & 0.63 \\
$V_{\rm H^{13}CO^{+},sys}^{\rm Dendro}$ $^\dag$ [km s$^{-1}$]   & 11.48        & 11.37        & 11.09        & 11.33 \\
$M_{\rm VIR}^{\rm Dendro}$ $^\ddag$ $[M_{\rm \odot}]$           & 0.66         & 1.81         & 1.03         & 1.72 \\
$\alpha_{\rm VIR}^{\rm Dendro}$ $^*$                            & 0.6          & 0.6          & 2.3          & 0.7 \\
\hline
\end{tabular}
\begin{tablenotes}
\item[$^{**}$] RA$_{\rm J2000}^{\rm Herschel}$ and DEC$_{\rm J2000}^{\rm Herschel}$ are the positions of cores identified by \citet{Konyves19}. RA$_{\rm J2000}^{\rm Dendro}$ and DEC$_{\rm J2000}^{\rm Dendro}$ are the positions of cores identified by Dendrogram.
\item[$^\dag$] $dV_{\rm H^{13}CO^{+},FWHM}^{\rm Herschel}$, $V_{\rm H^{13}CO^{+},sys}^{\rm Herschel}$, $dV_{\rm H^{13}CO^{+}, FWHM}^{\rm Dendro}$, and $V_{\rm H^{13}CO^{+},sys}^{\rm Dendro}$ are measured from the NOEMA+45m H$^{13}$CO$^+$ (1-0) map. $dV_{\rm H^{13}CO^{+},FWHM}^{\rm Herschel}$ are measured from the spectrum averaged toward the area with a center of RA$_{\rm J2000}^{\rm Herschel}$ and DEC$_{\rm J2000}^{\rm Herschel}$ and a radius of $R_{\rm core}^{\rm Herschel}$ from the NOEMA+45m H$^{13}$CO$^+$ (1-0) map. $dV_{\rm H^{13}CO^{+}, FWHM}^{\rm Dendro}$ are measured from the spectrum averaged toward the area identified as the corresponding structure from the Nobeyama+45m H$^{13}$CO$^+$ (1-0) map. 
\item[$^\ddag$] $M_{\rm VIR}^{\rm Herschel} = 125 R_{\rm core}^{\rm Herschel} (dV_{\rm H^{13}CO^{+},FWHM}^{\rm Herschel})^2$ and $M_{\rm VIR}^{\rm Dendro} = 125 R_{\rm core}^{\rm Dendro} (dV_{\rm H^{13}CO^{+},FWHM}^{\rm Dendro})^2$.
\item[$^*$] $\alpha_{\rm VIR}^{\rm Dendro}=M_{\rm VIR}^{\rm Dendro}/M_{\rm core}^{\rm Dendro}$ and $\alpha_{\rm VIR}^{\rm Herschel}=M_{\rm VIR}^{\rm Herschel}/M_{\rm core}^{\rm Herschel}$.
\item[$^+$] From \citet{Konyves19}. 
\item[$^{++}$] This object is identified as Class-0/I object [MGM]2882 by \citet{Megeath12}, while it is identified as prestellar core in \citet{Konyves19}. The separation between HGBS 054149.5-015941 and [MGM]2882 is 7$\arcsec$.2 which is larger than the threshold of 6$\arcsec$ of the cross-matching adopted in \citet{Konyves19}.
\end{tablenotes}
\end{threeparttable}
\end{table*}

%Table4
%\input{table4.tex}
\begin{table*}
\centering
\begin{threeparttable}
\caption{H$^{13}$CO$^+$ cores identified by {\it Dendrogram} in the $NOEMA$$+$45m data cube \label{table_core_noema}}
\begin{tabular}{lcccccc}
\hline
ID & 
RA$_{\rm J2000}$ &
DEC$_{\rm J2000}$ &
$R_{\rm core}^{\rm NOEMA+45m}$ & 
$V_{\rm sys}^{\rm NOEMA+45m}$ &
$dV_{\rm FWHM}^{\rm NOEMA+45m}$ & 
$M_{\rm VIR}^{\rm NOEMA+45m}$ \\
 & 
 &
 &
[pc] & 
[km s$^{-1}$] &
[km s$^{-1}$]  & 
[$M_{\odot}$]  \\
\hline
1  & 05$^{\rm h}$41$^{\rm m}$57$^{\rm s}$.1 & -02$^{\circ}$01$^{\rm m}$00$^{\rm s}$.0 & 0.020   & 11.53 & 0.51 & 0.63 \\
2  & 05$^{\rm h}$41$^{\rm m}$51$^{\rm s}$.4 & -01$^{\circ}$59$^{\rm m}$58$^{\rm s}$.1 & 0.013   & 11.50 & 0.51 & 0.42 \\
3  & 05$^{\rm h}$41$^{\rm m}$52$^{\rm s}$.8 & -02$^{\circ}$00$^{\rm m}$08$^{\rm s}$.7 & 0.010   & 11.45 & 0.45 & 0.26 \\
4  & 05$^{\rm h}$41$^{\rm m}$51$^{\rm s}$.7 & -01$^{\circ}$59$^{\rm m}$33$^{\rm s}$.6 & 0.008   & 11.38 & 0.42 & 0.18 \\
5  & 05$^{\rm h}$41$^{\rm m}$53$^{\rm s}$.7 & -02$^{\circ}$00$^{\rm m}$47$^{\rm s}$.7 & 0.007   & 11.11 & 0.74 & 0.51 \\
6  & 05$^{\rm h}$41$^{\rm m}$53$^{\rm s}$.5 & -02$^{\circ}$00$^{\rm m}$22$^{\rm s}$.6 & 0.012   & 11.32 & 0.47 & 0.32 \\
7  & 05$^{\rm h}$41$^{\rm m}$49$^{\rm s}$.4 & -01$^{\circ}$59$^{\rm m}$11$^{\rm s}$.7 & 0.008   & 11.32 & 0.64 & 0.42 \\
8  & 05$^{\rm h}$41$^{\rm m}$49$^{\rm s}$.7 & -01$^{\circ}$59$^{\rm m}$44$^{\rm s}$.6 & 0.007   & 11.24 & 0.52 & 0.23 \\
9  & 05$^{\rm h}$41$^{\rm m}$50$^{\rm s}$.2 & -01$^{\circ}$59$^{\rm m}$20$^{\rm s}$.8 & 0.006   & 11.18 & 0.66 & 0.32 \\
10 & 05$^{\rm h}$41$^{\rm m}$50$^{\rm s}$.3 & -02$^{\circ}$00$^{\rm m}$22$^{\rm s}$.4 & 0.011   & 10.96 & 0.55 & 0.43 \\
11 & 05$^{\rm h}$41$^{\rm m}$48$^{\rm s}$.9 & -02$^{\circ}$00$^{\rm m}$06$^{\rm s}$.3 & 0.008   & 11.03 & 0.57 & 0.30 \\
12 & 05$^{\rm h}$41$^{\rm m}$52$^{\rm s}$.1 & -02$^{\circ}$00$^{\rm m}$36$^{\rm s}$.5 & 0.008   & 10.88 & 0.74 & 0.54 \\
\hline
\end{tabular}
\begin{tablenotes}
\item[$^\ddag$] $M_{\rm VIR}^{\rm NOEMA+45m} = 125 R_{\rm core}^{\rm NOEMA+45m} (dV_{\rm FWHM}^{\rm NOEMA+45m})^2$.
\end{tablenotes}
\end{threeparttable}
\end{table*}

To identify cores in the NOEMA$+$45m H$^{13}$CO$^+$ (1--0) data cube, we applied the dendrogram algorithm with {\it min$\_$value}=4$\sigma$, {\it min$\_$delta}=4$\sigma$, and {\it min$\_$npix}=14.9 pixels (=$A_{\theta_{\rm beam}}/A_{\rm pixel}$, where $A_{\theta_{\rm beam}}$ and $A_{\rm pixel}$ are the surface area of the beam and pixel). We note that {\it min$\_$npix} is the total number of pixels where the structure is detected overall velocity channels. Here, we used the signal-to-noise ratio map to avoid the detection of spurious sources due to a nonuniform noise distribution. After performing a dendrogram analysis with these parameters, we rejected ambiguous or fake core candidates which do not have {\it min$\_$npix} pixels in two or more contiguous velocity channels. 

To identify dendrogram cores in the $Herschel$ column density map, we applied the algorithm with $A_{\rm V}$ = 8\footnote{Here, we focus on the identification of cores in the NGC~2024S filament. Prestellar cores are typically found in filaments with $A_{\rm V}$ values above 8 in {\it Herschel} data \citep{Konyves15}. Thus, we adopted \hbox{$A_{\rm V} = 8$} for {\it min$\_$value}.} \citep[assuming $N_{\rm H_2}/A_{\rm V} = 0.94 \times 10^{21} {\rm cm}^{-2}$,][]{Bohlin78} for the {\it min$\_$value}, $A_{\rm V}$ = 1 for {\it min$\_$delta}, and 28.9~pixels (=$A_{\theta_{\rm beam}}/A_{\rm pixel}$) for {\it min$\_$npix}.   

In this way, we identified twelve cores in the NOEMA+45m H$^{13}$CO$^+$ data cube. We also extracted four cores in the portion of the $Herschel$ column density map covered by NOEMA (which has an effective resolution $\theta_{\rm beam}$=18$\arcsec$.2$\sim$0.04 pc, see Fig. \ref{fig9}a). The positions of the cores identified here are consistent with those found in the $Herschel$ data by \citet{Konyves19}. As can be seen in Fig. \ref{fig4}c, each core detected in the $Herschel$ map corresponds to a single core in the NOEMA+45m H$^{13}$CO$^+$ data, suggesting that the $Herschel$ cores do not have significant substructure at a scale of $\sim$5$\arcsec$ ($\sim$ 0.01 pc or 2000 au). The positions of the cores identified in the $Herschel$ and NOEMA+45m data are listed in Tables \ref{table_core} and \ref{table_core_noema}, respectively.

The mass of each dendrogram core was estimated as 

%Eq 1
\begin{equation}
M_{\rm core}^{\rm Dendro} [M_{\odot}] = N({\rm H}_2) m_{\rm H} \mu_{\rm H_2} A_{\rm core}^{\rm Dendro},
\end{equation}

\noindent where $m_{\rm H}$ is the hydrogen atom mass, $\mu_{\rm H_2}$ = 2.8 is the mean molecular weight per H$_{2}$ molecule, and $A_{\rm core}^{\rm Dendro}$ is the projected area of each core identified by the dendrogram analysis. Here, the total $N({\rm H}_2)$ was measured using the Bijection scheme as defined by  \citet{Rosolowsky08}. The uncertainty in $M_{\rm core}^{\rm Dendro}$ is typically a factor of 2, mainly due to uncertainties in the dust opacity \citep[cf.][]{Roy14}. The core masses from the dendrogram analysis range from 0.44 $M_{\odot}$ to 2.84 $M_{\odot}$, with a mean value $<$$M_{\rm core}^{\rm Dendro}$$>$\,=\,1.7$\pm$1.0 $M_{\odot}$. The core masses reported by \citet{Konyves19} have a mean value $<$$M_{\rm core}^{\rm Herschel}$$>$\,=\,2.5$\pm$1.2 $M_{\odot}$ (see also Table \ref{table_core}) and are consistent within better than a factor of $\sim$2 with the masses from the dendrogram analysis. The main reason why the two sets of mass estimates differ slightly is  that the dendrogram technique does not subtract background emission and returns different source sizes.

Under the assumption that each core has a spheroidal shape and a density profile of $\rho \varpropto r^{-2}$, we also estimated the virial masses $M_{\rm VIR}^{\rm Herschel}$ and $M_{\rm VIR}^{\rm Dendro}$ of the detected cores as follows \citep[see][]{Ikeda07, Shimajiri15a},

%Figure 6
\begin{figure*}
\centering
\includegraphics[angle=0,width=9cm]{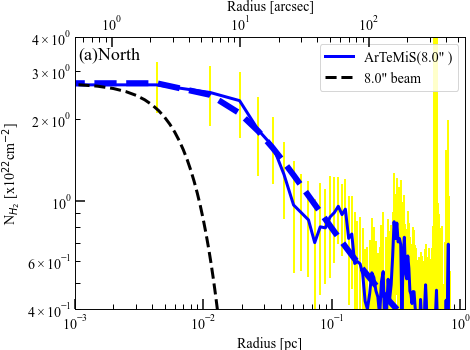}
\includegraphics[angle=0,width=9cm]{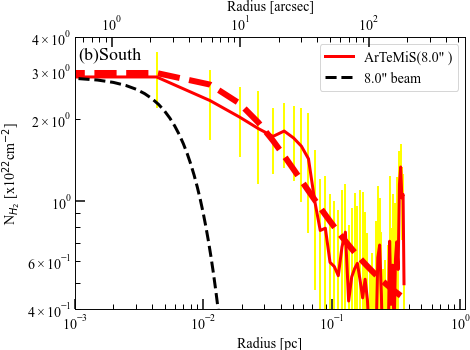}
\caption{Median radial ArT$\acute{e}$MiS+$Herschel$ column density profiles for the (a) northeastern and (b) southwestern side of the NGC~2024S filament. The defined crest of the filament is shown in Fig. \ref{figB1}. The black dashed curves indicate the angular resolution of the ArT$\acute{e}$MiS+$Herschel$ column density map (8$\arcsec$). The dashed curves show the best-fit Plummer mode. The yellow bars show the dispersion ($\pm$1$\sigma$) of the distribution of radial profiles along the filament. The area affected by the secondary component seen in the NOEMA H$^{13}$CO$^+$ data was avoided when producing the median radial profile for the southwestern side of the NGC~2024S filament (see Sect. \ref{sect:high-reso_map}). 
}
\label{fig6}
\end{figure*}

%Eq 2
\begin{equation}
M_{\rm VIR}^{\rm Herschel} [M_{\odot}]
= \frac{3 R_{\rm core}^{\rm Herschel} \sigma_{\rm Herschel}^2 }{ G } 
= 125 R_{\rm core}^{\rm Herschel} \left(dV^{\rm Herschel}_{\rm H^{13}CO^{+}, FWHM}\right
)^2
\end{equation}

or 

%Eq 3
\begin{equation}
M_{\rm VIR}^{\rm Dendro} [M_{\odot}]
= \frac{3 R_{\rm core}^{\rm Dendro} \sigma_{\rm Dendro}^2 }{ G } 
= 125 R_{\rm core}^{\rm Dendro} \left(dV^{\rm Dendro}_{\rm H^{13}CO^{+}, FWHM}\right
)^2.
\end{equation}

\noindent The radius $R_{\rm core}^{\rm Herschel}$ is provided in \citet{Konyves19}. The radius $R_{\rm core}^{\rm Dendro}$ of each core was estimated as $R_{\rm core}^{\rm Dendro}$ [pc] = $\sqrt{A_{\rm core}^{\rm Dendro}/\pi}$. The velocity dispersion $\sigma_{\rm Herschel}$ and $\sigma_{\rm Dendro}$ was determined as $\sigma_{\rm Herschel} = dV^{\rm Herschel}_{\rm H^{13}CO^{+},FWHM} / \sqrt{8\ln 2}$ and $\sigma_{\rm Dendro} = dV^{\rm Dendro}_{\rm H^{13}CO^{+},FWHM} / \sqrt{8\ln 2}$, where $dV^{\rm Dendro}_{\rm H^{13}CO^{+},FWHM}$ is the mean FWHM velocity width of the Nobeyama+45m H$^{13}$CO$^{+}$(1--0) emission among pixels in $A_{\rm core}^{\rm Dendro}$ and $dV^{\rm Herschel}_{\rm H^{13}CO^{+},FWHM}$ is the mean FWHM velocity width of the Nobeyama+45m H$^{13}$CO$^{+}$(1--0) emission among the area withing a $R_{\rm core}^{\rm Herschel}$ from the core position (RA$_{\rm J2000}^{\rm Herschel}$, DEC$_{\rm J2000}^{\rm Herschel}$). The mean FWHM velocity width among cores ranges from 0.4 km s$^{-1}$ to 0.7 km s$^{-1}$, while the typical FWHM velocity along the filamentary structure is 0.6 km s$^{-1}$. The virial mass ratios $\alpha_{\rm VIR}^{\rm Herschel} (\equiv M_{\rm VIR}^{\rm Herschel}/M_{\rm core}^{\rm Herschel})$ and $\alpha_{\rm VIR}^{\rm Dendro} (\equiv M_{\rm VIR}^{\rm Dendro}/M_{\rm core}^{\rm Dendro})$  are lower than $\sim$2, suggesting that all four cores are gravitationally bound. The derived physical parameters of each core are given in Table \ref{table_core}.

%%%%%%%%%%%%%%%%%%%%%%%%%%%%%%%%%%%%%%%%%%
% Section 4.1: Filament properties
%%%%%%%%%%%%%%%%%%%%%%%%%%%%%%%%%%%%%%%%%%
\subsubsection{Filament properties}\label{sect. fil_pro}

Figure \ref{fig6} shows, in log-log format, the median radial column density profiles measured on the northeastern and southwestern sides of the NGC~2024S filament in the 8$\arcsec$-resolution ArT$\acute{\rm e}$MiS+$Herschel$ data whose background emission is not subtracted. Here, the filament crest was defined using the DisPerSE algorithm \citep{Sousbie11,Sousbie11b, Arzoumanian11}. Following \citet{Arzoumanian11} and \citet{Palmeirim13}, we fitted the density profiles on the northeastern and southwestern sides of the NGC~2024S filament with a Plummer-like model as below: 

%Eq. 4
\begin{equation}
N_{\rm H_2} (r) = \frac{N_{\rm H_2}^0}{\biggl[1+(r/R_{\rm flat})^2\biggr]^{\frac{p-1}{2}}} + {\rm B_{kg}},
\end{equation}

\noindent where $N_{\rm H_2}^0$ is the column density at filament center, $R_{\rm flat}$ is the radius of the flat inner region, $p$ is the power-law exponent at larger radii, and B$_{\rm kg}$ is the column density of the background. $N_{\rm H_2}^0$ is expressed as $A_{\rm p} \rho_{\rm c} R_{\rm flat}/\mu m_{\rm H}$ where $A_{\rm p}$ = $\frac{1}{\cos i}$ $\times$ B$\Bigl(\frac{1}{2},\frac{p-1}{2} \Bigr)$ is a finite constant factor. The factor $\frac{1}{\cos i}$ takes into account the inclination of the filament to the plane of the sky. Here, we assumed $i$=0. For a population of randomly oriented filaments with respect to the plane of the sky, the net effect is a factor of <$\frac{1}{\cos i}$>$\sim$ 1.57 on average \citep[cf.][]{Arzoumanian11}. B$\Bigl(\frac{1}{2},\frac{p-1}{2} \Bigr)$ is the Euler beta function. The fitting results are summarized in Table~\ref{table4}. The density at the center of the filament is estimated to be $n_{\rm c}$=$(1.2\pm0.4)\times$10$^5$ cm$^{-3}$ from the Plummer fits to the radial profile averaged between the southwestern and northeastern sides of the filament.

The half-power diameter of the filament as derived from Plummer fitting, $D_{\rm HP}^{\rm Plummer}$, corresponds to:

%Eq. 5
\begin{equation}
D_{\rm HP}^{\rm Plummer} = \sqrt{2^{\frac{2}{p-1}}-1} \times D_{\rm flat},
\end{equation}

\noindent where $D_{\rm flat} \equiv 2 \times R_{\rm flat}$. $D_{\rm HP}^{\rm Plummer}$ provides a more robust measurement of the inner width of a Plummer-like profile than $D_{\rm flat}$ since its derivation is not as strongly correlated to that of $p$ \citep[cf.][]{Schuller21}. The half-power diameter $D_{\rm HP}^{\rm Plummer}$ of the NGC~2024S filament as derived from fitting the northeastern and southwestern sides of the median radial profile simultaneously is {0.081$\pm$0.014} pc. (The $D_{\rm HP}^{\rm Plummer}$ values obtained by fitting the northeastern and southwestern sides of the radial profile separately are {0.078$\pm$0.015} pc and {0.071$\pm$0.040} pc, respectively.)\footnote{Note that the $p$ values differ slightly for each of the fit (see Table~\ref{table4}), which explains why the $D_{\rm HP}^{\rm Plummer}$ value from the two-sided fit is not a simple average of the two values obtained from the one-sided fits.}. These values agree well with the half-power widths found in $Herschel$ studies of Gould Belt filaments \citep{Arzoumanian11, Arzoumanian19, Palmeirim13}.

We also estimate the virial mass of the filament, $M_{\rm line, vir}$ $\equiv$ $2$ $\sigma^2/G$ $\sim$ 84 $(\frac{dV_{\rm FWHM}}{{\rm km\ s}^{-1}})^2$ [$M_{\odot} {\rm pc}^{-1}$] \citep[][]{Fiege00}. The mean velocity width, $dV_{\rm FWHM}$, is measured to be 0.62 km s$^{-1}$ (min:0.49 km s$^{-1}$, max:0.84 km s$^{-1}$) from the Nobeyama H$^{13}$CO$^+$ (1--0) map. Note that the velocity width is measured toward the whole area mapped by the NOEMA. Thus, the virial mass of the NGC~2024S is 32.4 $M_{\odot} {\rm pc}^{-1}$ (min: 20.2 $M_{\odot} {\rm pc}^{-1}$, max:59.0 $M_{\odot} {\rm pc}^{-1}$).

%Table 5
%\input{table5.tex}
\begin{table*}
\centering
\begin{threeparttable}
\caption{Properties of the NGC 2024S filament \label{table4}}
\begin{tabular}{lcccc}
\hline
Parameter & Northeastern & Southwestern & Averaged  & B211/B213\\
          & side$^\dag$  &  side$^{\dag,*}$ & both sides & filament$^\ddag$\\
  (1)    & (2)                &  (3)                     & (4)             &(5).                       \\          
\hline
$N_{\rm H_2}^0$ [$\times$10$^{21}$ cm$^{-2}$] & 26.1$\pm$0.2    & 27.2$\pm$0.4    & 26.2$\pm$0.2     & 14.4$\pm$1.4\\
$R_{\rm flat}$  [pc]                          & 0.019$\pm$0.005 & 0.021$\pm$0.011 & 0.021$\pm$0.005  & 0.032$\pm$0.014 \\
$p$                                           & 1.8$\pm$0.2     & 2.0$\pm$0.6     & 1.9$\pm$0.2      & 2.0$\pm$0.09 \\
$D^{\rm Plummer}_{\rm HP}$  [pc]              & 0.078$\pm$0.015	& 0.071$\pm$0.040 &	0.081$\pm$0.014$^{\star}$  & 0.11$\pm$0.02 \\
$B_{\rm kg}$  [$\times$10$^{21}$ cm$^{-2}$]   & 1.58$\pm$0.07   & 2.98$\pm$2.42   & 1.56$\pm$0.01	 & 0.67$\pm$0.17 \\
$n_{\rm c}$ [$\times$10$^{5}$ cm$^{-3}$]      & 1.2$\pm$0.5     & 1.4$\pm$0.1     & 1.2$\pm$0.4      & 0.45 \\
\hline
\end{tabular}
\begin{tablenotes}
\item[\dag] Fitting results on the 8$\arcsec$ resolution ArT$\acute{\rm e}$MiS+{\it Herschel} column density map.
\item[*] The area affected by the secondary component seen in NOEMA H$^{13}$CO$^+$ is avoided to produce the median radial profile (see Sect. \ref{sect:high-reso_map}). 
\item[\ddag] \citet{Palmeirim13}.
\item[$\star$] 
$D^{\rm Plummer}_{\rm HP}$ value obtained by fitting the northeastern and southwestern sides of the median radial profile simultaneously. 
Due to differing $p$ indices, 
this is not a simple average of the $D^{\rm Plummer}_{\rm HP}$ values given in columns [2] and [3].
\end{tablenotes}
\end{threeparttable}
\end{table*}

%Fig. 7
\begin{figure*}
\centering
\includegraphics[angle=0,width=18cm]{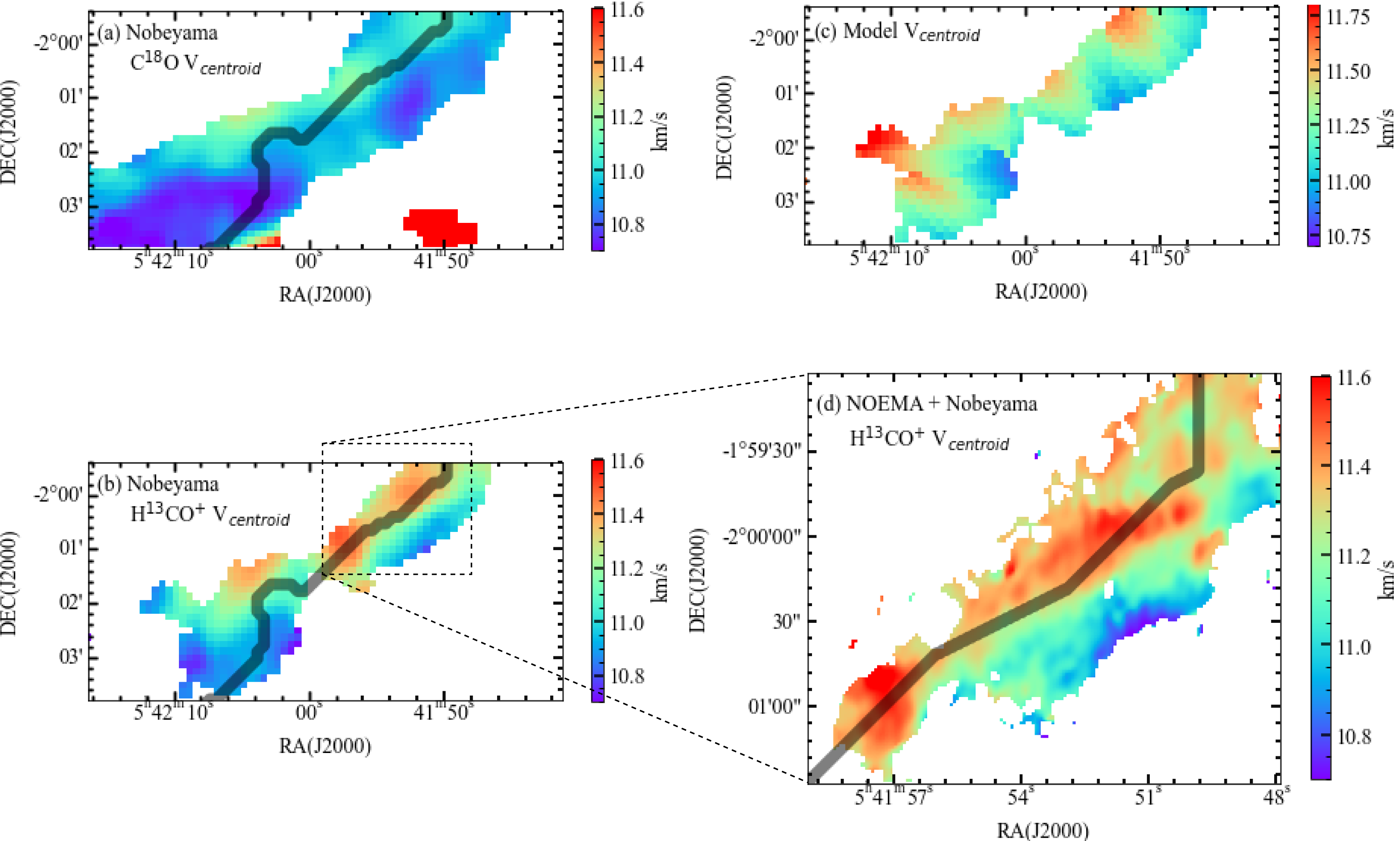}
\caption{($a$) Nobeyama C$^{18}$O (1--0) centroid velocity map, ($b$) Nobeyama H$^{13}$CO$^+$(1--0) centroid velocity map (close-up view of $panel$ b in Fig. \ref{fig4}), ($c$) centroid velocity map of the fragmenting filament model, and ($d$) NOEMA+45m H$^{13}$CO$^+$(1--0) centroid velocity map toward the NOEMA observed area indicated by the dashed box in panel b. In $panels$ a and b, a black line indicates the crest of the filament which corresponds to the $z$-axis in Fig. \ref{fig8}a and $r$=0 in Fig. \ref{fig8}b. The back line starts at (RA$_{\rm J2000}$,DEC$_{\rm J2000}$)=(5$^{\rm h}$42$^{\rm m}$11$^{\rm s}$.823, -2$^{\circ}$02$^{\rm m}$53$^{\rm s}$.97) and ends at (RA$_{\rm J2000}$,DEC$_{\rm J2000}$)=(5$^{\rm h}$41$^{\rm m}$45$^{\rm s}$.806, -1$^{\circ}$59$^{\rm m}$23$^{\rm s}$.97). See also Fig. \ref{figA1}.
}
\label{fig7}
\end{figure*}

\subsection{Velocity structure}\label{sect:velocity_structure}

The spectra in the Nobeyama C$^{18}$O (1--0) and H$^{13}$CO$^+$ (1--0) data cubes show a single velocity component at all positions. Thus, we performed a Gaussian fitting analysis with a single component for all spectra. In this way, we obtained the centroid velocity at each pixel in the Nobeyama C$^{18}$O (1--0)  and H$^{13}$CO$^+$ (1--0) data. Figure~\ref{fig7} shows the C$^{18}$O (1--0) and H$^{13}$CO$^+$ (1--0) centroid velocity maps. In the southeastern part of the dense elongated ridge corresponding to NGC~2024, a velocity gradient  can be seen along the  direction perpendicular to the filament. In the present paper, we refer to the southeastern part of the filamentary structure in NGC~2024, indicated by a red box in Fig. \ref{fig4}b, as the NGC~2024S filament. In this area, the blue- and red-shifted H$^{13}$CO$^{+}$ velocity components are distributed on the southwestern and the northeastern side of the filament crest, respectively, indicating the presence of a transverse velocity gradient across the filament.

%Table 6
%\input{table6.tex}
\begin{table*}
\tiny
\centering
\caption{Positional offsets between column density and velocity peaks \label{table_offset}}
\scalebox{0.76}{
\begin{tabular}{|l|c|c|cccc|}
\hline
                         & \multirow{2}{*}{$\lambda$} &       &   HGBS    &  HGBS& HGBS  & HGBS  \\
                         &   &       &    054203.2-02035    &   054157.3-020101 &  054153.4-020016  &  054149.5-015941 \\
\hline
{\it Herschel} column density             & ----      & peak  & $z$=0.48~pc                  &  $z$=0.81~pc             & $z$=0.96~pc              & $z$=1.10~pc \\ 
\hline
\multirow{3}{*}{Nobeyama H$^{13}$CO$^+$}  & \multirow{3}{*}{0.21~pc}   & peak(obs./fit)          &  $z$=0.59~pc/0.61~pc    &  $z$=0.81~pc/0.81~pc    &   $z$=1.02~pc/1.02~pc & --- \\ 
                         &                            & offset(obs./fit) in pc          & $-$0.11~pc/$-$0.13~pc                  &  0~pc/0~pc             & $-$0.062~pc/$-$0.062~pc              & --- \\ 
                         &                            & offset(obs./fit) in $\lambda$   & $-$0.54$\pm$0.12$\lambda$/$-$0.62$\pm$0.12$\lambda$  & 0.00$\pm$0.12$\lambda$/0.00$\pm$0.12$\lambda$        & $-$0.30$\pm$0.12$\lambda$/$-$0.30$\pm$0.12$\lambda$     & --- \\ 
                         &                            & offset(obs./fit) in $\lambda$ $^{\dag}$  & $+$0.06$\pm$0.12$\lambda$/$-$0.005$\pm$0.12$\lambda$  & --- & --- & --- \\ 
\hline
\multirow{3}{*}{Nobeyama C$^{18}$O}  & \multirow{3}{*}{0.26~pc}   & peak(obs./fit)               &   $z$=0.61~pc/0.59~pc           &   $z$=0.83~pc/0.85~pc    & ----   &   $z$=1.13~pc/1.11~pc  \\ 
                         &                            & offset(obs./fit) in pc          &   $-$0.13~pc/$-$0.11~pc              &  $-$0.020~pc/$-$0.040~pc             & ----             &  $-$0.029~pc/$-$0.038~pc\\ 
                         &                            & offset(obs./fit) in $\lambda$   &   $-$0.50$\pm$0.09$\lambda$/$-$0.42$\pm$0.09$\lambda$  &   $-$0.07$\pm$0.09$\lambda$/$-$0.15$\pm$0.09$\lambda$             & ----     & $-$0.11$\pm$0.09$\lambda$/$-$0.16$\pm$0.09$\lambda$ \\ 
                         &                            & offset(obs./fit) in $\lambda$ $^{\dag}$  & $+$0.05$\pm$0.09$\lambda$/$-$0.10$\pm$0.09$\lambda$  & --- & --- & --- \\ 

\hline
\end{tabular}
}
\begin{tablenotes}
\item Note: $z$ measures position along the filament crest shown in Fig. \ref{fig1}.
The uncertainties of the positional offsets in units of $\lambda$ were estimated assuming that the peaks in column density have an uncertainty corresponding to half a beam. 
\item Note$^{\dag}$: Given the additional errors arising from the definition 
of the curved filament crest around HGBS 054203.2-02035, we also provide offsets estimated assuming 
a straight filament crest toward this core (see text in Sect.~\ref{sect:Jeans}).
\end{tablenotes}
\end{table*}

%Note for Table 7 (straight crest)
%Npeak 0.242 pc
%H13CO+
% obs 0.25442302 pc --> +0.0124 pc --> 0.06 lambda +/- 0.12 lambda
% fit 0.241 pc --? -0.001 pc --> -0.005 lambda +/- 0.12 lambda
%
%C18O 
% obs 0.25442302 pc --> +0.0124 pc --> +0.05 lambda +/- 0.09 lambda
% fit 0.217 pc --> -0.025 pc --> - 0.10 lambda +/- 0.09 lambda

\begin{figure*}
\centering
\includegraphics[angle=0,width=17.5cm]{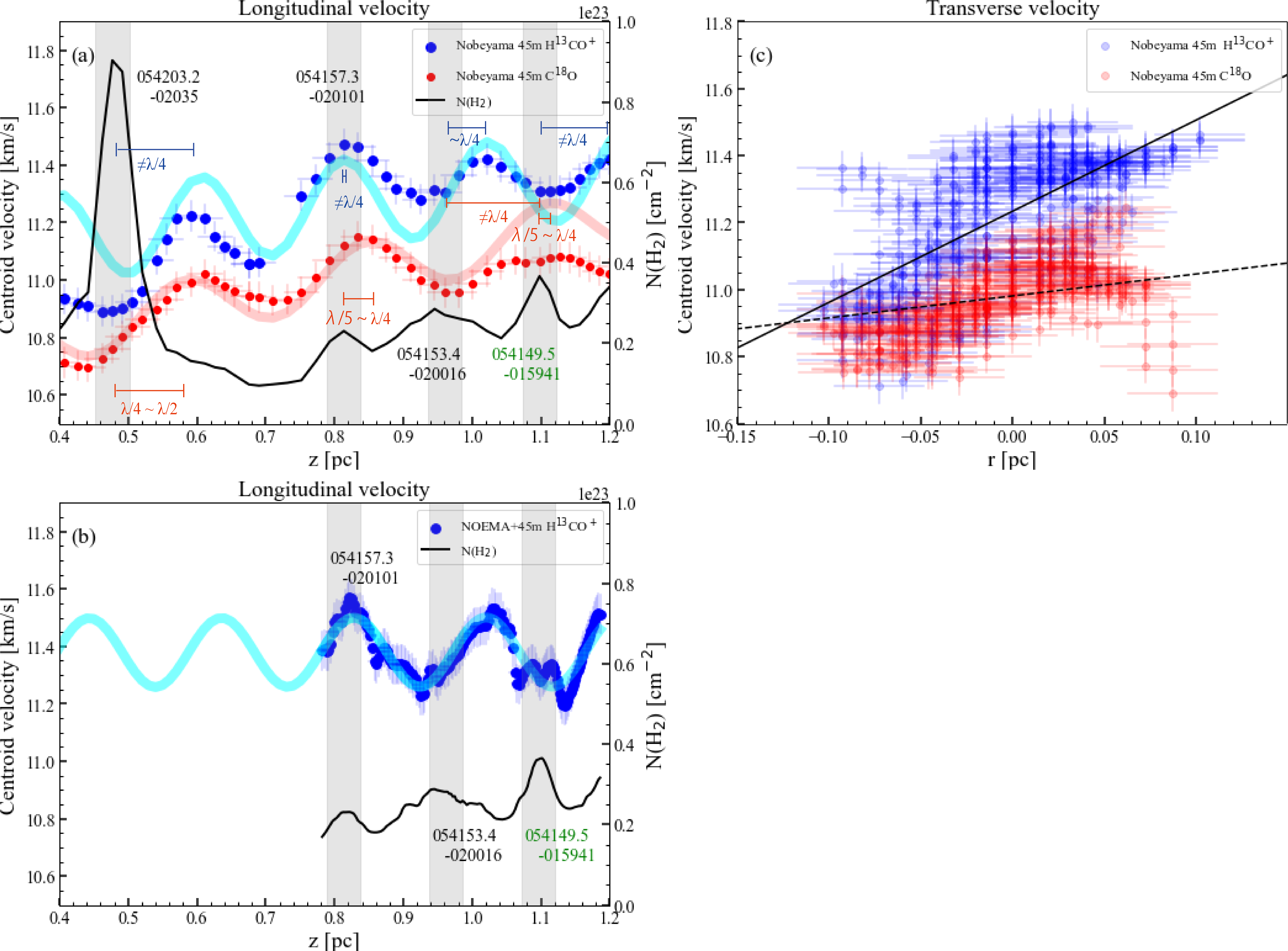}
\caption{($a$) Nobeyama 45m H$^{13}$CO$^{+}$ (1--0) and C$^{18}$O (1--0) centroid velocities along the filament, ($b$) NOEMA+45m H$^{13}$CO$^{+}$ (1--0) centroid velocities along the filament, and ($c$) Nobeyama 45m H$^{13}$CO$^{+}$ (1--0) and C$^{18}$O (1--0) centroid velocities along the $r$-direction. In each panel, blue and red points the centroid velocity of H$^{13}$CO$^{+}$ and C$^{18}$O, respectively. In $panel$ a and b, the blue and red curves shows the result of the least square fitting with a function of $v(z) = V_{\rm sys} +  z \nabla V_z  + V_0 \cos(2\pi z/\lambda +\theta_{\rm offset})$ against the H$^{13}$CO$^{+}$ (1--0) and C$^{18}$O(1--0) centroid velocity. In $panel$ a and b, the black curves indicate the distribution of {\it Herschel} H$_2$ column density along the filament in the 25$\arcsec$ resolution map. The vertical grey strips indicate the positions of $Herschel$ dense cores identified by \citet{Konyves19} (HGBS~054203.2-020235, 054157.3-020101, 054153.4-020016, and 054149.5-015941). The width of each strip corresponds to a 25$\arcsec$ beam. The core labeled in green is associated with a {\it Spitzer} protostar \citep{Megeath12}. 
In this plot, $z$ measures position along the magenta curve in Fig. \ref{fig1} and $z$=0 corresponds to the southeastern edge of the curve. 
Each data point is on the crest of the filament. In $panel$ c, the two solid and dashed lines show the best-fit transverse velocity gradient of the form $V(r) =V_{\rm sys} +  r \nabla V_r$ observed in H$^{13}$CO$^{+}$ (1--0) and C$^{18}$O (1--0), respectively. $r$=0 corresponds to the crest of the filament as indicated by the magenta curve in Fig. \ref{fig1}. All pixels in the maps of Fig. \ref{fig7} (a, b, and d) are used for this plot by estimating the projected separation from the filament crest.
}
\label{fig8}
\end{figure*}

Figure \ref{fig8}a shows the variations in Nobeyama H$^{13}$CO$^+$ and C$^{18}$O centroid velocities along the NGC~2024S filament ($z$-axis). A velocity oscillation pattern can be recognized along the filament. Using the following fitting function\footnote[1]{For the fitting, we used the python scipy.optimize.curve\_fit package
} \citep[see][]{Peretto16}:
%(\url{https://docs.scipy.org/doc/scipy/reference/generated/scipy.optimize.curve_fit.html})} \citep[see][]{Peretto16}:

%Eq 1
\begin{equation}
V(z) = V_{\rm sys} +  z \nabla V_z  + V_0 \cos(2\pi z/\lambda +\theta_{\rm offset} ), 
\end{equation}

\noindent the best-fit velocity gradient along the filament $\nabla V_z$ was found to be 0.29$\pm$0.17 km s$^{-1}$ pc$^{-1}$ in H$^{13}$CO$^+$ and 0.51$\pm$0.06 km s$^{-1}$ pc$^{-1}$ in C$^{18}$O. The amplitude ($V_{0}$) of the oscillation was found to be 0.15$\pm$0.08 km s$^{-1}$ in H$^{13}$CO$^+$ and 0.10$\pm$0.02 km s$^{-1}$ in C$^{18}$O, respectively. The wavelength ($\lambda$) of the oscillation pattern was found to be 0.21$\pm$0.01~pc in H$^{13}$CO$^+$ and 0.26$\pm$0.01~pc in C$^{18}$O, respectively. The positions of the dense cores identified by \citet{Konyves19} the $Herschel$ GBS data are indicated by vertical grey strips at $z$ = 0.48~pc (HGBS 054203.2-02035), $z$ = 0.81~pc (HGBS054157.3-020101), $z$ = 0.96 pc (HGBS 054153.4-020016), and $z$ = 1.10~pc (HGBS 054149.5-015941). They are slightly shifted from the observed centroid velocity peaks in H$^{13}$CO$^+$ or C$^{18}$O. 
Table \ref{table_offset} summarizes the offsets between the {\it Herschel} column density peaks 
and the Nobeyama 45m H$^{13}$CO$^+$ and C$^{18}$O peaks. Here, $z$ measures position along the filament crest shown in Fig. \ref{fig1} and 
$z$ = 0~pc corresponds to southeastern edge of the filament crest. The positional offset between the peak in column density and that in centroid velocity for HGBS 054203.2-02035 is about $-0.1$~pc, roughly corresponding to $-0.5\pm0.1\, \lambda$ in H$^{13}$CO$^+$ and about $-0.13$~pc, roughly corresponding to $-0.5\pm0.1\, \lambda$ in C$^{18}$O. The positional offset for HGBS 054157.3-020101 is $-0.020$~pc in the observed velocity pattern and $-0.04$~pc in the fitted velocity pattern, roughly corresponding to $\sim -0.1--0.2 \lambda$, in C$^{18}$O, while the {\it Herschel} dense core located at 0.81~pc coincides with the peak in H$^{13}$CO$^+$ centroid velocity. The positional difference for HGBS 054153.4-020016 is $-0.062$~pc, roughly corresponding to $-1/4\,\lambda$, in H$^{13}$CO$^+$, while the C$^{18}$O emission does not show a clear peak in centroid velocity. The positional difference for HGBS 054149.5-015941 is $-0.03$~pc in the observed velocity pattern, roughly corresponding to $-0.1\, \lambda$ and $-0.04$~pc in the observed velocity pattern corresponding to $\sim -0.2 \lambda$ in C$^{18}$O, while the H$^{13}$CO$^+$ emission does not show clear peak in centroid velocity. This source is associated with a {\it Spitzer} protostar in the catalog of \citet{Megeath12}.

Figure \ref{fig8}b shows the variations in NOEMA+45m H$^{13}$CO$^+$ centroid velocity along the NGC~2024S filament ($z$-axis). The distribution of the NOEMA+45m H$^{13}$CO$^+$ centroid velocity is consistent with that of the Nobeyama H$^{13}$CO$^+$ centroid velocity. Only three cores are covered because of the limited extent of the NOEMA observations. However, the velocity pattern seen in the NOEMA+45m H$^{13}$CO$^+$ data is more nicely fitted than the Nobeyama H$^{13}$CO$^+$ pattern.

%Figure 8

Figure \ref{fig8}c shows the variations in H$^{13}$CO$^+$ centroid velocity along the minor axis of the filament ($r$-direction), confirming the presence of a transverse velocity gradient (i.e., the centroid velocity is redshifted to the northeast of the filament crest, while it is blueshifted to the southwest of the crest). We note that this velocity gradient has a direction opposite to that seen on larger (parsec) scales as described in Sect.~\ref{sec:results-CO} (see Fig. \ref{fig3}). Indeed, the maps observed in CO (and isotopes) show emission at redshifted velocities to the southwest of the filament (Figs. \ref{fig2}d, i, n) and blueshifted velocities to the northeast of the filament (Figs. \ref{fig2}b, g, l). In addition, the dimensionless coefficient $C_{v} \equiv \frac{\delta V^2}{GM_{\rm line}}$ introduced by \citet{Chen20},  where $\delta V$ is half of the velocity difference across the filament, is estimated to be much less than 1 ($C_{v}$=0.12), suggesting that the transverse velocity gradient observed in H$^{13}$CO$^+$ on small scales is driven by self-gravity as opposed to large-scale shock compression. The H$^{13}$CO$^+$ velocity gradient may reflect bulk motion of the filament itself (e.g., such as possibly rotation of the filament about its main axis) \citep[cf.,][]{Matsumoto94,Dhabal18,Hsieh21}, although this would require confirmation.

For simplicity, we fitted the centroid velocities observed along the minor axis of the filament assuming a constant transverse velocity gradient $\nabla V_r$, as in the following equation:

%Eq 6
\begin{equation}
V(r) = V_{\rm sys} + r \nabla V_r.
\end{equation}

\noindent The best-fit transverse velocity gradient (in the $r$-direction) is found to be 2.72$\pm$0.15 km s$^{-1}$ pc$^{-1}$. It is worth noting that this transverse velocity gradient is an order of magnitude higher than the longitudinal velocity gradient of 0.29$\pm$0.17 km s$^{-1}$ pc$^{-1}$. The fit parameters are summarized in Table \ref{table3}. For comparison, the transverse velocity gradients observed in the Orion~A integral-shaped filament and in the SDC13 infrared dark filament are measured to be $\sim$1.0 km s$^{-1}$ in H$^{13}$CO$^+$ (1--0) and 0.2--1.5 km s$^{-1}$ pc$^{-1}$ in NH$_3$ (1,1) \citep{Ikeda07, Williams18}, respectively.

%Table 7
%\input{table7.tex}
\begin{table*}
\centering
\begin{threeparttable}
\caption{Fitting results for the distribution of observed centroid velocities \label{table3}}
\begin{tabular}{lcc}
\hline
Line                            & H$^{13}$CO$^+$(1--0) & C$^{18}$O(1--0) \\
\hline
Transverse velocity gradient       & \multicolumn{2}{c}{($V(r) = V_{\rm sys} + r \nabla V_r$)} \\
$V_{\rm sys}$                   & 11.23$\pm$0.01 km s$^{-1}$            & 10.98$\pm$0.01  km s$^{-1}$\\
$\nabla$$V_r$                   & 2.72$\pm$0.15 km s$^{-1}$ pc$^{-1}$   & 0.65$\pm$0.19  km s$^{-1}$ pc$^{-1}$ \\
\hline
Longitudinal velocity gradient     & \multicolumn{2}{c}{($V(z) = V_{\rm sys} +  z \nabla V_z  + V_0 \cos(2\pi z/\lambda +\theta_{\rm shift})$)} \\
$V_{\rm sys}$                   & 11.03$\pm$0.14 km s$^{-1}$            & 10.64$\pm$0.05 km s$^{-1}$ \\
$\nabla V_z$                    & 0.29$\pm$0.17 km s$^{-1}$ pc$^{-1}$   & 0.51$\pm$0.06 km s$^{-1}$ pc$^{-1}$ \\
$V_0$                           & 0.15$\pm$0.08 km s$^{-1}$             & 0.10$\pm$0.02 km s$^{-1}$ \\
$\lambda$                       & 0.21$\pm$0.01 pc                      & 0.26$\pm$0.01  pc \\
$\theta_{\rm offset}$           & 0.55$\pm$1.46 rad                      & 4.90$\pm$0.64  rad \\
\hline
\end{tabular}
\begin{tablenotes}
\item Note: The fitting was performed on independent centroid velocity measurements (i.e., separated by more than a beam size).  Quoted errors are formal statistical errors corresponding to the
square roots of the diagonal elements in the covariance matrix of fitted parameters 
returned by the {\it scipy curve\_fit} routine. Total errors may be larger.
\end{tablenotes}
\end{threeparttable}
\end{table*}

Figure \ref{fig8} also shows the variations in C$^{18}$O centroid velocity along (panel $a$) and across  (panel $c$) the filament. In both directions, the C$^{18}$O centroid velocities differ somewhat  the centroid velocities observed in H$^{13}$CO$^+$. Possible reasons why the C$^{18}$O and H$^{13}$CO$^+$ centroid velocities exhibit slightly different patterns are that i) the C$^{18}$O molecule may be depleted in the inner part of the filament as described in Sect. \ref{sec:results-centroid}, and ii) the C$^{18}$O emission preferentially traces the outer parts of the filament compared to the H$^{13}$CO$^+$ emission as described in Sect. \ref{sect. fil_pro}.

We also derived a velocity structure function (VSF) $S_{2}(\mbox{\boldmath $l$})$ from the Nobeyama H$^{13}$CO$^+$ data. The function $S_{2}(\mbox{\boldmath $l$})$ at each scale $l$ can be defined as follows \citep[cf.][]{Peretto16, Henshaw20}: 

%Eq 8
\begin{equation}\label{Eq:Sv}
S_{2}(\mbox{\boldmath $l$}) = {\rm median} \left( [V(x_i,y_i+\mbox{\boldmath $l$}) - V(x_i,y_i)]^2 \right),
\end{equation}

\noindent where $(x_{i},y_{i})$ are the coordinates of each position and $l$ denotes the separation between positions.

\begin{figure*}
\centering
\includegraphics[angle=0,width=16cm]{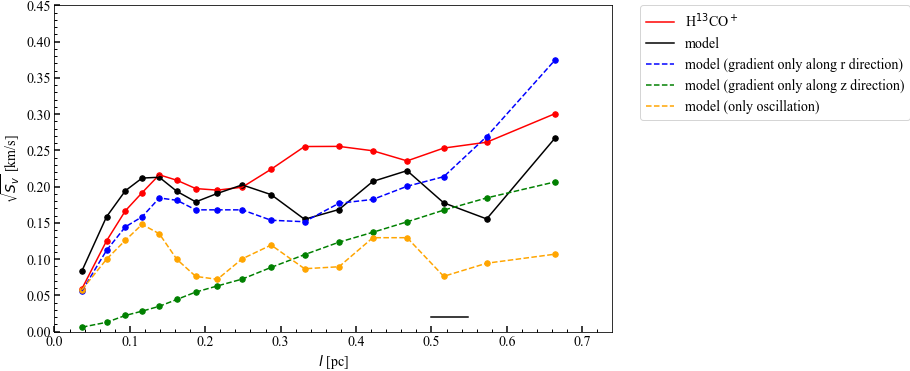}
\caption{Comparison of the velocity structure function between Nobeyama H$^{13}$CO$^+$ (1--0) data and the model. Red line and points indicate the velocity structure function of the Nobeyama H$^{13}$CO$^+$ (1--0) centroid velocity. Black line and points indicate the velocity structure function of the modeled fragmenting filament. Blue, green, and blue dashed lines indicate the VSF of the models which are taken into account only velocity gradient along $r$-direction, velocity gradient along $z$-direction, and oscillation, respectively. The typical uncertainty of the observed VSF is 0.08 km s$^{-1}$. 
}
\label{fig9}
\end{figure*}

Figure~\ref{fig9} shows the observed VSF of the Nobeyama H$^{13}$CO$^+$ velocity components. The VSF increases up to 0.2 km s$^{-1}$ for $\mbox{\boldmath $l$}$ $<$ $\sim$0.15 pc, then increases more gradually with small oscillations for $\sim$0.15 pc $\le$ $\mbox{\boldmath $l$}$ $<$ $\sim$0.6 pc, and finally increases steadily again for$\sim$0.6 pc $\le$ $\mbox{\boldmath $l$}$. 

%Figure 9

%%%%%%%%%%%%%%%%%%%%%%%%%%%%%%%%%%%%%%%%%%
% Section 4 :Discussion
%%%%%%%%%%%%%%%%%%%%%%%%%%%%%%%%%%%%%%%%%%
\section{Discussion}\label{Sect4}

\subsection{Variations in filament width among tracers}\label{Sect41}

%Figure 10
\begin{figure}
\centering
\includegraphics[angle=0,width=8cm]{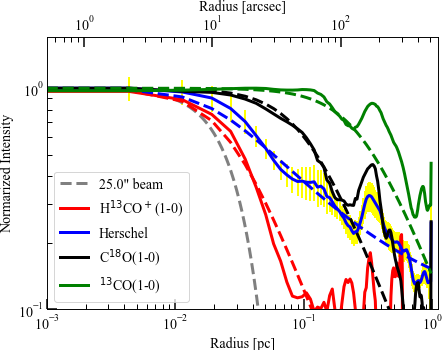}
\caption{Comparison of median radial column density profiles for the northeastern side of the NGC~2024S filament among (blue) $Herschel$ column density, (red) H$^{13}$CO$^{+}$, (black) C$^{18}$O, and (green) $^{13}$CO at a resolution of 25$\arcsec$ ($\sim$0.048 pc). The dashed curves show the best-fit Plummer model. The yellow bars show the dispersion ($\pm$1$\sigma$) of the distribution of the radial profile along the filament in $Herschel$. The grey curves indicate the angular resolution of 25$\arcsec$. Note that we reproduced the Nobeyama H$^{13}$CO$^+$ map with an angular resolution of 25$\arcsec$ to compare it with others in the same angular resolution.}
\label{fig10}
\end{figure}

%Table 8
%\input{table8.tex}
\begin{table*}
\centering
\begin{threeparttable}
\caption{Comparison of width estimates among tracers for the NGC~2024S filament \label{table_filament_width}}
\begin{tabular}{lcccc}
\hline

Tracer                          &   beam size [$\arcsec$]               &    beam size [pc]                 & $D_{\rm HP}^{\rm Plummer}$  [pc] $^{*}$ & $p$   \\
\hline
ArT$\acute{e}$MiS+$Herschel$    & 8$\arcsec$                            & $\sim$0.015                    & 0.081$\pm$0.014                 & 1.9$\pm$0.2   \\
$Herschel$                      & 18$\arcsec$.2                         & $\sim$0.035                     & 0.080$\pm$0.010                 & 1.7$\pm$0.1   \\
$Herschel$                      & 25$\arcsec$.0                         & $\sim$0.048                     & 0.097$\pm$0.012                 & 1.7$\pm$0.1   \\
H$^{13}$CO$^+$(1--0)$^\dag$     & 6$\arcsec$.40$\times$3$\arcsec$.68    & $\sim$0.011  $\times$ 0.007   & 0.047$\pm$0.005                 & 2.5$\pm$0.2  \\
H$^{13}$CO$^+$(1--0)            & 25$\arcsec$.0                         & $\sim$0.048                   & 0.063$\pm$0.012                 & 2.5$\pm$0.4   \\
C$^{18}$O(1--0)                 & 25$\arcsec$.0                         & $\sim$0.048                     & 0.251$\pm$0.021                 & 2.5$\pm$0.2   \\
$^{13}$CO(1--0)                 & 25$\arcsec$.0                         & $\sim$0.048                     & 0.694$\pm$0.485                 & 2.5$\pm$1.0    \\
\hline
\end{tabular}
\begin{tablenotes}
\item[$^{*}$] Deconvolved from the beam size. 
\item[$^\dag$] NOEMA+45m H$^{13}$CO$^+$(1--0) integrated intensity map.
\end{tablenotes}
\end{threeparttable}
\end{table*}

As described in Sect. \ref{Sect3}, filamentary structures are detected in the Nobeyama $^{13}$CO (1--0), C$^{18}$O (1--0), and H$^{13}$CO$^+$ (1--0) data. \citet{Panopoulou14} measured the widths of $^{13}$CO (1--0)  filamentary structures in the Taurus molecular cloud and found a typical value of 0.4 pc, while \citet{Palmeirim13} found a filament width of $\sim$0.1 pc from the $Herschel$ column density map of the Taurus B211/213 region. Using N$_{2}$H$^+$ (1--0) and H$^{13}$CO$^+$ (1--0) intensity maps of the Serpens Main, Perseus, and Orion~A molecular clouds, \citet{Lee14}, \citet{Dhabal18}, and \citet{Hacar18} reported a typical filament width of $\sim$0.035 pc, which is narrower than the value of $\sim 0.1$~pc found for $Herschel$ filaments \citep{Arzoumanian11, Arzoumanian19}. In order to investigate whether these differences in filament width estimates arise from using different tracers, we fitted the integrated intensity profiles observed in $^{13}$CO (1--0), C$^{18}$O (1--0), H$^{13}$CO$^+$ (1--0) on the northeastern side of the NGC~2024S filament in the same manner as in Sect. \ref{sect. fil_pro} for the column density profiles: 

%Eq. 9
\begin{equation}
W_{\rm mol}(r) = \frac{W_{\rm mol}^0}{\biggl[1+(r/R_{\rm flat})^2\biggr]^{\frac{p-1}{2}}},
\end{equation}

\noindent where $W_{\rm mol}(r)$ is the integrated intensity of each observed molecular transition. To compare the widths of the filament  obtained from the $^{13}$CO (1--0), C$^{18}$O (1--0), $Herschel$ column density, and H$^{13}$CO$^{+}$ (1--0) maps, we fitted the data at the same angular resolution of 25$\arcsec$. We found $D_{\rm HP}^{\rm Plummer}$ values of 0.694$\pm$0.485 pc in $^{13}$CO, 0.251$\pm$0.021 pc in C$^{18}$O, 0.097$\pm$0.012 pc with {\it Herschel}, and 0.063$\pm$0.012 pc in H$^{13}$CO$^+$, respectively (see also Fig. \ref{fig10} and Table \ref{table_filament_width}). The measured $D_{\rm HP}^{\rm Plummer}$ width is only marginally resolved in the Nobeyama H$^{13}$CO$^+$ (1--0) data at 25$\arcsec$ resolution ($\sim$0.048 pc). Therefore, we also fitted the NOEMA+45m H$^{13}$CO$^+$ data at an angular resolution of $\sim$6$\arcsec$.4$\times$3$\arcsec$.7 (0.012 pc $\times$ 0.006 pc). We found a $D_{\rm HP}^{\rm Plummer}$ value of 0.047$\pm$0.005 pc, which is a factor of 2 lower than the $D_{\rm HP}^{\rm Plummer}$ width measured in the $Herschel$ column density map. The $^{13}$CO (1--0), C$^{18}$O (1--0), and H$^{13}$CO$^{+}$ (1--0) data trace emission regions of density $\sim$10$^{3}$ cm$^{-3}$, $\sim$10$^{3-4}$ cm$^{-3}$, $\sim$10$^{3-5}$ cm$^{-3}$, respectively \citep{Onishi98,Yonekura05,Ikeda07,Maruta10,Qian12,Shimajiri15a}. Our results for the NGC~2024S filament (see, e.g., Fig.~\ref{fig10}) confirm that filament widths measured in dense gas tracers such as N$_2$H$^+$ (1--0) and H$^{13}$CO$^+$ (1--0) tend to be narrower than those found using tracers of low-density gas such as $^{13}$CO (1--0) and C$^{18}$O (1--0) \citep{Panopoulou14,Lee14,Dhabal18,Hacar18}. The observed differences in filament width measurements among tracers are likely due to differences in the range of densities probed by each tracer. The Nobeyama $^{13}$CO (1--0) and C$^{18}$O (1--0) data trace the outer (lower density) part of the $Herschel$ filament and the Nobeyama/NOEMA H$^{13}$CO$^+$ (1--0) emission trace the inner (denser) part. This also shows that it is important to compare measurements obtained with the same tracer when discussing the universality (or non-universality) of filament widths. The filament profiles obtained in any given molecular line tracer are affected by a limited dynamic range in density, as described above,  and are sensitive to chemical effects such as depletion \citep{Bergin02, Tafalla02} or far-ultraviolet (FUV) photo-dissociation \citep{Lada94,Shimajiri14,Lin16}. Using N(H$_2$) column density profiles derived from high dynamic range submm dust continuum maps (from, e.g., {\it Herschel}) provides more reliable estimates of filament widths.

The $D_{\rm HP}^{\rm Plummer}$ width we measure here in C$^{18}$O (1--0) for the NGC~2024S filament is 0.2 pc, while a more ``typical'' filament width of $\sim$0.12$\pm$0.04pc (FWHM) was reported by \citet{Orkisz19} based on Gaussian fitting for a sample of C$^{18}$O (1--0) filaments observed with the IRAM 30m telescope in Orion~B. In the Orkisz et al. study, the FWHM widths of filaments in the NGC~2024 subregion tend to be broader than the ``typical'' value in the sample and reach up to 0.2 pc. Thus, our C$^{18}$O (1--0) findings for NGC~2024S are consistent with the results of Orkisz et al.

It is also worth comparing the half-power diameter of the NGC~2024S filament in Orion~B with the filament widths found in the Orion~A molecular cloud. Recently, \citet[][]{Schuller21} measured the distribution of filament half-power diameters in the northern part of the Integral shaped filament (ISF) of the Orion~A molecular cloud using  APEX/ArT\'eMiS 350 and 450$\mu$m data combined with {\it Herschel}/SPIRE data, providing an angular resolution of 8$\arcsec$ (corresponding to 0.015 pc at a distance of 410 pc, \citealp{Menten07}). They found that half-power diameters ranging from 0.06 pc to 0.11 pc and line masses in the range $\sim$100--500 $M_{\odot}$/pc. The half-power diameters of the massive star-forming filament in NGC~6334 ($M_{\rm line}$ = 600--1200 $M_{\odot}$ pc$^{-1}$ rescaled to a distance of 1.35 kpc, \citealp{Chibueze14}) and the low-mass star-forming filament B211/B213 in Taurus ($M_{\rm line}$ = 54 $M_{\odot}$ pc$^{-1}$) have been measured to be $D_{\rm HP}^{\rm Plummer}$=0.12$\pm$0.03pc (at $d$ = 1.35 kpc) and $D_{\rm HP}^{\rm Plummer}$=0.11$\pm$0.02pc, respectively. The half-power diameter of 0.081$\pm$0.014 pc reported here for the NGC~2024S filament ($M_{\rm line}$ = 62 $M_{\odot}$ pc$^{-1}$) is consistent with the values found  in the ISF. We conclude that the half-power diameters measured from submm dust continuum data are consistent among filaments spanning a wide range of line masses.

%%%%%%%%%%%%%%%%%%%%%%%%%%%%%%%%%%%%%%%%%%%%%%%%%%%%%%%%%%%%%%%
% Section 4 : Filament fragmentation into cores
%%%%%%%%%%%%%%%%%%%%%%%%%%%%%%%%%%%%%%%%%%%%%%%%%%%%%%%%%%%%%%%
\subsection{Filament fragmentation}

As described in Sect.~\ref{sect:velocity_structure}, a positional offset is seen between the peak in H$^{13}$CO$^+$(1--0) integrated intensity and the peak in H$^{13}$CO$^+$(1--0) centroid velocity. Furthermore, the centroid velocity observed along the NGC~2024S filament  exhibits an oscillation pattern. A similar velocity structure was reported by \citet{Hacar11} for a filament in the Taurus/L~1517 region. In L~1517, a $\lambda$/4 phase shift was observed between the density and the velocity field around the cores forming along the filament axis. \citet{Hacar11} argued that the L~1517 filament was in the process of fragmenting owing to gravitational instability. Here, we similarly discuss whether the NGC~2024S filament may be fragmenting into cores due to gravitational instability based on a comparison between the density and the velocity field (Sect. \ref{sect:Jeans}), and a comparison of the observed velocity structure function with that of a toy model of a fragmenting filament (Sect. \ref{Sect:VSF}).

%%%%%%%%%%%%%%%%%%%%%%%%%%%%%%%%%%%%%%%%%%
% Section 4.2 : Filament properties
%%%%%%%%%%%%%%%%%%%%%%%%%%%%%%%%%%%%%%%%%%
%Figure 8
\begin{figure}
\centering
\includegraphics[angle=0,width=9cm]{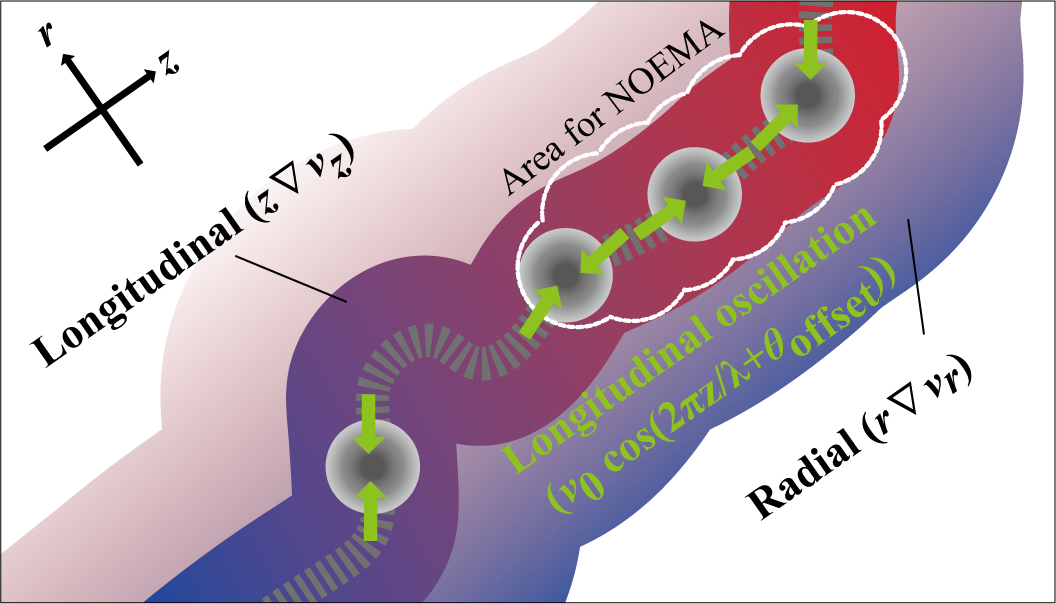}
\caption{Schematic picture of the velocity structure in NGC~2024S based on our observational results. Darker red and blue colors indicate the velocity gradient in longitudinal direction. Lighter red and blue colors indicate the velocity gradient in the radial direction. A dashed gray line marks the crest of the filament. Filled gray circles indicate the cores. Green arrows indicate the velocity oscillation along the filament. A white polygon indicates the area observed with NOEMA. }
\label{fig11}
\end{figure}

\subsubsection{Fragmentation by gravitational instability?}\label{sect:Jeans}

As discussed in Sect.~\ref{sect:velocity_structure} (see, e.g., Fig. \ref{fig8}a), a positional offset is observed between the column density peaks and the peaks in either or both H$^{13}$CO$^+$ or/and C$^{18}$O centroid velocity. A $\lambda$/4 phase shift between the density and the velocity field is expected for core-forming motions in a filament fragmenting into condensations \citep[cf.][]{Gehman96, Hacar11}. For cores to be forming, gas motions have to converge into the core centers. Therefore, the density peak associated with a forming core has to correspond to a position of vanishing velocity. This requires a $\lambda$/4 phase shift between the density and the centroid velocity under the assumption that the density and velocity perturbations are sinusoidal. The condensation seen in the $Herschel$ column density map at $z$ = 0.96~pc (HGBS 054153.4-020016) shows a $\lambda$/4 phase shift between the density and the H$^{13}$CO$^+$ centroid velocity and corresponds to a protostar identified by \citet{Megeath12}.  This supports the view that the observed velocity and column density patterns trace the convergence of matter onto the corresponding protostellar core. However, a clear C$^{18}$O velocity peak associated with HGBS 054153.4-020016 is not observed. In the case of the HGBS 054157.3-020101 condensation, the column density peak coincides with the H$^{13}$CO$^+$ velocity peak, while a $\sim \lambda$/5-$\lambda$/4 phase shift is observed between the column density and C$^{18}$O velocity peaks. For HGBS 054149.5-015941, a $\sim\lambda$/5-$\lambda$/4 phase shift between the column density and C$^{18}$O velocity peaks is also observed but is not seen in H$^{13}$CO$^+$. For HGBS 054203.2-02035, the phase shift is almost $\lambda$/2 in both H$^{13}$CO$^+$ and C$^{18}$O. However, the precise location of the filament crest around HGBS~054203.2-02035 is uncertain due, e.g., to the lower column density of the filament between HGBS~054203.2-02035 and HGBS~054157.3-020101. While the nominal crest orientation around HGBS~054203.2-02035 is almost south to north, the overall crest orientation of the NGC~2024S filament is from southeast to northwest. The actual filament crest around HGBS~054203.2-02035 is believed to be located somewhere in between the south-north and southeast-northwest directions. Assuming a straight crest from southeast to northwest, the column density and velocity peaks almost coincide in both H$^{13}$CO$^+$ and C$^{18}$O (i.e., $\sim$ 0 $\lambda$ shift). Therefore, given the uncertainty in the exact location of the filament crest, an actual shift $\sim \lambda$/4 cannot be ruled out for HGBS~054203.2-02035. To summarize, a $\sim \lambda$/4 shift is observed around HGBS 054153.4-020016 in H$^{13}$CO$^+$ and the data around HGBS 054203.2-02035, 054157.3-02010, and 054149.5-015941 are marginally consistent with $\sim \lambda$/4 shifts in C$^{18}$O given the error bars, but there is no evidence of a $\sim \lambda$/4 shift in H$^{13}$CO$^+$ around HGBS 054203.2-02035, 054157.3-02010, 054149.5-015941, nor in C$^{18}$O around HGBS 054153.4-020016. The difference between H$^{13}$CO$^+$ and C$^{18}$O patterns may arise from differences in the range of densities probed by H$^{13}$CO$^+$ and C$^{18}$O as discussed in Sect. \ref{Sect41}.

In the quasistatic fragmentation model of \citet{Gehman96}, a systematic phase shift of $\lambda$/4 between the density and velocity peaks is expected for all cores. Here, a $\lambda$/4 shift  is not observed for all cores in any single tracer although it is tentatively observed for all cores in either H$^{13}$CO$^+$ or C$^{18}$O. Thus, the physical structure of the NGC~2024S filament is clearly more complex than the prediction of the simple quasistatic fragmentation model. Strictly speaking, the quasistatic model discussed by \citet{Gehman96} is only expected to apply to isolated, nearly isothermal filaments close to hydrostatic equilibrium. In the case of the NGC~2024S filament, both the line mass ($M_{\rm line}$ $\sim$ 62 $M_{\odot}$ ${\rm pc}^{-1}$) and the index of the radial density profile ($p\sim 2$) differ from the thermally critical line mass $M_{\rm line, crit} \equiv 2\,c_s^2/G \sim$16 $M_{\odot}\, {\rm pc}^{-1}$ and $p=4$ value expected for an isothermal filament in hydrostatic equilibrium \citep[cf.][]{Ostriker64}. The NGC~2024S filament is nevertheless close to virial equilibrium with $M_{\rm line} \sim M_{\rm line, vir} \equiv 2\,\sigma^2/G $ \citep[][see also Sect. \ref{sect. fil_pro}]{Fiege00}. Moreover, both polytropic and magnetized equilibrium filaments may have $p\sim 2$ as observed for NGC~2024S \citep[cf.][]{Kawashi1998,Palmeirim13,Kashiwagi21}. A more important difference perhaps with the idealized quasistatic model of \citep{Gehman96} is that the NGC~2024S filament is not isolated but embedded in the turbulent environment of the Orion~B cloud and may be accreting from this environment. As illustrated by the numerical simulations of \citet{Clarke16} and \citet{Anathpindika21}, the ambient environment may modify the fragmentation properties of a filament. 

%%%%%%%%%%%%%%%%%%%%%%%%%%%%%%%%%%%%%%%%%%%%%%%%%%%%%%%%%%%%%%%
% Section 4.3: Velocity structure function
%%%%%%%%%%%%%%%%%%%%%%%%%%%%%%%%%%%%%%%%%%%%%%%%%%%%%%%%%%%%%%%

\subsubsection{Modeling the fragmenting filament}\label{Sect:VSF}

To examine whether the observed velocity pattern can be explained by filament fragmentation, we modeled the velocity field in and around the filament taking into account four velocity components
\citep[cf.][]{Peretto16}:

%Eq 10
\begin{equation}\label{eq:model}
V(z,r) = V_{\rm sys} + r \nabla V_r + z \nabla V_z  + V_0 \cos(2\pi z/\lambda+\theta_{\rm offset})
\end{equation}

\noindent  The first term on the right hand side of Eq. (\ref {eq:model}) expresses the systemic velocity of the cloud. The second and third terms express transverse and longitudinal velocity gradients, respectively. The fourth term expresses a longitudinal oscillation caused by fragmentation of the filament into cores. $z$ and $r$ denote the longitudinal (major axis of the filament) and the radial (minor axis of the filament) direction, respectively. $\nabla V_r$ and $\nabla V_z$ are the velocity gradients along the radial and longitudinal directions, respectively. $V_0$ and $\lambda$ are the amplitude and wavelength of the longitudinal oscillations. The values of $\nabla V_z$, $\nabla V_r$, $V_0$, and $\lambda$ were obtained from the fitting results (see Sect.~\ref{sect:velocity_structure}, Table \ref{table3}, and Fig.~\ref{fig8}). Figure~\ref{fig11} shows a schematic picture of the velocity structure in NGC~2024S based on our observational results. Figures~\ref{fig7}c and \ref{fig7}d show the centroid velocity map of the toy model and that of the Nobeyama H$^{13}$CO$^+$ (1--0) data, respectively. It can be seen that the observed distribution of H$^{13}$CO$^+$ centroid velocities is similar to that of the model.

To get further insight into the physical meaning of the observed velocity structure (see Sect. \ref{sect:velocity_structure}), we compared the observed velocity structure function to the velocity structure function obtained from our toy model, as well as single-component models taking each velocity component separately into account.

Figure \ref{fig9} compares the observed VSF from the Nobeyama H$^{13}$CO$^+$ data cube with the VSF of the model including the above three velocity components [black curve -- see Eq. (\ref{eq:model})]. The VSF of the model, $[S_{2}(\mbox{\boldmath $l$})]^{1/2}$, increases up to 0.2 km s$^{-1}$ for $\mbox{\boldmath $l$}$ $<$ $\sim$0.15 pc, then increases more gradually with small oscillations for $\sim$0.15 pc $\le$ $\mbox{\boldmath $l$}$ $<$ $\sim$0.6 pc, and finally increases steadily again for$\sim$0.6 pc $\le$ $\mbox{\boldmath $l$}$. We also show in Fig.~\ref{fig9} the VSFs of each model component separately: i) the transverse velocity component (blue curve), ii) the longitudinal velocity component (green curve), and iii) the longitudinal oscillation component caused by fragmentation (yellow curve). Among the single-component model VSFs, the only VSF showing an oscillation pattern is that of the model with a longitudinal oscillation. The observed H$^{13}$CO$^+$ VSF does show an oscillation pattern and is qualitatively very similar to the VSF of our  model including  all three velocity components (see Fig.~\ref{fig9}). This suggests that the oscillation pattern seen in the observed velocity structure function results from the effect of gravitational fragmentation of the NGC2024S filament into cores. \citet{Hacar16} performed a similar velocity structure function analysis using $^{13}$CO (2--1) data toward the 6-pc long filament in the Musca cloud and found that the observed VSF could be described by the superposition of a global velocity gradient along the filament and local velocity oscillations.

\subsubsection{Core separation along the filament}

When its mass per unit length is close to that required for hydrostatic equilibrium, a filament is expected to fragment into cores with a characteristic spacing of about 4 times the filament width according to the self-similar models which describe the evolution of isothermal filaments under the influence of self-gravity without magnetic fields or turbulence \citep{Inutsuka92}.

As described in Sect.~\ref{sect. fil_pro}, the filament diameter $D_{\rm HP}^{\rm Plummer}$ of NGC~2024S is estimated to be $\sim$0.081$\pm$0.014 pc. Thus, the typical separation between cores is expected to be $\sim$0.32 pc, corresponding to 4 times the observed filament width. Five cores (HGBS~054157.3-020101, 054153.4-020016, 054150.5-020024, 054149.5-015941, and 054203.2-020235) are embedded along the NGC~2024S filament. HGBS~054203.2-020235 is not covered by the NOEMA observations and HGBS~054150.5-020024 may be  associated with the secondary component seen in NOEMA H$^{13}$CO$^+$ as mentioned in Sect.~\ref{sect:high-reso_map}. The mean separation among the four $Herschel$ cores, excluding HGBS~054150.5-020024, is 0.12$\pm$0.05 pc. The mean separation among the five $Herschel$ cores is 0.13$\pm$0.06 pc. These are projected separations which do not take the viewing angle of the filament into account. Assuming the inclination of the filament to the line of sight is $\alpha_0 = 18\pm 5$ deg, the observed separations would translate into intrinsic separations consistent with $\sim$4 times the filament width. However, this would require the NGC~2024S filament to be seen closer to a ``pole-on''  configuration than to a ``plane-of-sky'' configuration. Assuming random orientations, the probability of observing a filament with a viewing angle $\alpha \leq \alpha_0 $ is $p = 1 - \rm{cos}\, \alpha_0 \sim$\,5\% for $\alpha_0 = 18$~deg. Alternatively, adopting a more likely inclination to the line of sight [e.g., $\alpha_0 \geq 60$ deg, for which $p(\alpha \leq \alpha_0) \geq 50\% $], the observed separations would be indicative of an intrinsic core spacing $\la$\,0.16~pc, significantly shorter than the separation predicted by the standard model of filament fragmentation. A similar trend is observed in several other filaments \citep[e.g.,][]{Tafalla15,Zhang20}. 

To test whether the observed separation among cores may be present in the case of randomly distributed cores, we conducted a total of 10000 realizations of random distributions of 5 sources in a 0.85-pc-long filament (see Fig.\ref{fig8}a) using the python code {\it FRAGMENT} \citep{Clarke19} and measured the separation among the randomly-placed sources. Comparison between the resulting overall distribution of nearest-neighbor separations (NNS) and the observed NNS distribution using a Kolmogorov-Smirnov (K-S) test yields a probability or ``p-value'' $p$=0.09 (equivalent to 1.6$\sigma$ in Gaussian statistics), indicating that the quasi-periodic pattern of the observed cores is only marginally significant. We therefore cannot rule out the possibility that the observed pattern arises from a random distribution.

\subsubsection{Relation between core mass and filament line mass}

\citet{Andre19} proposed that the prestellar core mass function (CMF) may be inherited from the filament mass function (FLMF) through gravitational fragmentation of individual filaments and suggested that higher-mass cores may form in higher $M_{\rm line}$ filaments. In their proposed empirical model, the mass of a core formed via fragmentation\footnote{In contrast to the idealized model presented by \citet{Gehman96}, recent observations suggest that the gravitational fragmentation of a quasi-equilibrium filament occurs in at least two stages or modes: ``cylindrical'' fragmentation leads to the formation of clumps along the filament, separated by $\sim 4$ times the filament width, and subsequent ``spherical'' fragmentation of the clumps generates Bonnor-Ebert-like cores within clumps \citep[e.g.][]{Kainulainen17,Shimajiri19b,Clarke17}. The average core mass is set by the effective critical Bonnor-Ebert mass in the clumps, which is itself related to the local surface density of the filament as per Eq.~\ref{eq:M_BE}.} of a thermally supercritical but virialized filament corresponds to the effective Bonnor-Ebert mass $M_{\rm BE, eff}$ in the filament (see \citealp{Andre19}):

%Eq 
\begin{equation}\label{eq:M_BE}
M_{\rm BE, eff} \sim 1.3 \frac{c_{\rm s,eff}^4}{G^2\Sigma_{\rm fil}},
\end{equation}

\noindent where $c_{\rm s,eff}$, $G$, and $\Sigma_{\rm fil}$ are the one-dimensional velocity dispersion or effective sound speed, the gravitational constant, and the surface density of the filament, respectively. Since the relation $M_{\rm line} \sim \Sigma_{\rm fil} \times D_{\rm HP}^{\rm Plummer} \sim M_{\rm line,vir} \equiv 2 c_{\rm s,eff}^2/G$ holds for a thermally supercritical filament \citep{Arzoumanian13}, we may expect the following relation between the typical core mass and the filament line mass:

\begin{equation}\label{eq:M_BE_eff}
\left(\frac{M_{\rm BE, eff}}{[M_{\odot}]}\right) \sim 0.325 \left(\frac{M_{\rm line}}{[M_{\odot}/{\rm pc}]}\right) \times \left(\frac{D_{\rm HP}^{\rm Plummer}}{[{\rm pc}]}\right).
\end{equation}

\noindent The $D_{\rm HP}^{\rm Plummer}$ width of the NGC~2024S filament is measured to be $\sim$0.081$\pm$0.014 pc. Thus, with $M_{\rm line} = 62\pm13\ M_{\odot} {\rm pc}^{-1}$, the core mass in the NGC~2024S filament is expected to be 1.6$\pm$0.4 $M_{\odot}$, which agrees very well with the observed mean core mass of 2.5$\pm$1.2 $M_{\odot}$.

\begin{figure}
\centering
\includegraphics[angle=0,width=8.5cm]{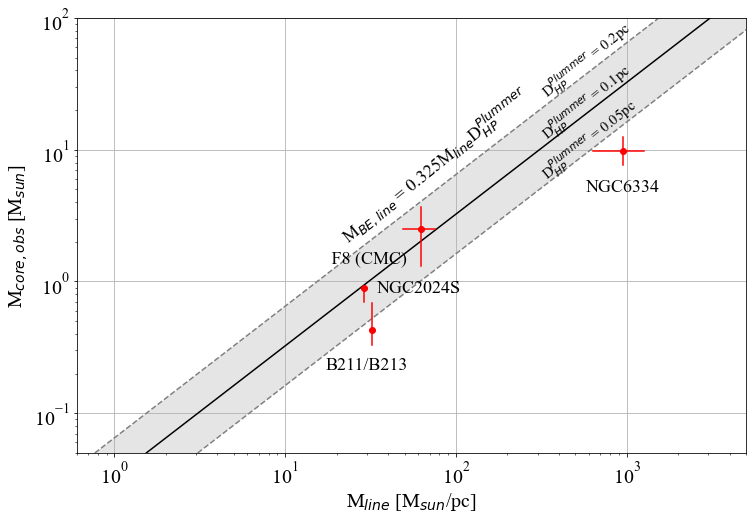}
\caption{$M_{\rm core,obs}$-$M_{\rm line}$ relation. The black solid line indicates $M_{\rm BE, eff} \sim 0.325 M_{\rm line} D_{\rm HP}^{\rm Plummer}$ where $D_{\rm HP}^{\rm Plummer}$=0.1pc. The black dashed line indicates $M_{\rm BE, eff} \sim 0.325 M_{\rm line} D_{\rm HP}^{\rm Plummer}$ where $D_{\rm HP}^{\rm Plummer}$=0.05 pc and 0.2 pc.
The $M_{\rm line}$, $M_{\rm core,obs}$, and their uncertainties for B211/B213, Filament 8 in CMC, NGC~6334 are from \citet{Marsh16}, \citet{Zhang20}, \citet{Andre16}, and \citet{Shimajiri19b}.}
\label{fig12}
\end{figure}

These findings can be compared to the results of other recent filament fragmentation studies. Our ALMA observations of the NGC~6334 filament \citep{Shimajiri19b} revealed 26 compact dense cores with a mean mass of 9.6$_{-1.9}^{+3.0}$ $M_{\odot}$ in this massive filament ($M_{\rm line}$ = 600--1200 $M_{\odot}$ pc$^{-1}$ rescaled to a distance of 1.35 kpc, \citealp{Chibueze14}). In their study of the X-shaped nebula in the California molecular cloud, \citet{Zhang20} identified cores with a mass of 0.9$^{+0.0}_{-0.2}$ $M_{\odot}$ within their Filament~8 ($M_{\rm line} \approx 30\, M_{\odot}$/pc). Therefore, we find that there is a suggestive trend of increasing $M_{\rm core,obs}$ with increasing $M_{\rm line}$ (see Fig.~\ref{fig12}), although more data points would be required to be conclusive. The observed $M_{\rm core,obs}$-$M_{\rm line}$ trend is roughly consistent with Eq.~(\ref{eq:M_BE_eff}), indicating that higher-mass cores may form in higher $M_{\rm line}$ filaments as proposed by \citet{Andre19} and \citet{Shimajiri19b}.

%fig 12
%\begin{comment}
%end{comment}

%%%%%%%%%%%%%%%%%%%%%
% Conclusions
%%%%%%%%%%%%%%%%%%%%%
\section{Conclusions}\label{Sect5}
To investigate the detailed velocity and density structure of a fragmenting filament in the NGC~2024 region of the Orion~B molecular cloud, we performed observations of the $^{12}$CO (1--0), $^{13}$CO (1--0), C$^{18}$O (1--0), and H$^{13}$CO$^+$ (1--0) molecular lines with the Nobeyama 45m telescope and the NOEMA interferometer. Our main results may be summarized as follows:

\begin{itemize}

\item We found that the Nobeyama $^{13}$CO (1--0), C$^{18}$O (1--0) and H$^{13}$CO$^+$ (1--0) emission traces the  filamentary structure that is seen in the $Herschel$ column density map.

\item Analysis of the median radial column density profiles of NGC~2024S from ArT$\acute{e}$MiS+$Herschel$ data yields an half-power diameter of $D_{\rm HP}^{\rm Plummer}$=$\sim$0.081$\pm$0.014 pc for the filament, which agrees well with the results of previous $Herschel$ filament studies in nearby molecular clouds. 

\item Comparison of the radial profiles derived from $Herschel$, Nobeyama H$^{13}$CO$^+$ (1--0), C$^{18}$O (1--0), and $^{13}$CO (1--0) data shows that measured filament widths may differ depending on the tracer used. Therefore, the same tracer must be employed to discuss the universality (or non-universality) of filament widths. As the filament profiles obtained in any given molecular line tracer are affected by a limited dynamic range in density, using N(H$_2$) column density profiles derived from, e.g., {\it Herschel} dust continuum maps provides more reliable estimates of filament widths.
            
\item Performing a dendrogram analysis, we detected twelve cores in the NOEMA+45m H$^{13}$CO$^+$ (1--0) map and four cores in the $Herschel$ column density map over the field observed with NOEMA. Each core detected in the $Herschel$ column density map 
corresponds to only one core detected by NOEMA, 
suggesting that the $Herschel$ cores do not have significant substructure. 

\item  The centroid velocity distribution along the major axis of the filament shows an oscillation pattern and a tentative $\lambda$/4 phase shift compared to the density distribution. This $\lambda$/4 shift is not simultaneously observed for all cores in any single tracer but is tentatively seen for each core in either H$^{13}$CO$^+$ or C$^{18}$O. The difference between the H$^{13}$CO$^+$ and C$^{18}$O velocity patterns may arise from differences in the range of densities probed by H$^{13}$CO$^+$ and C$^{18}$O. These results are consistent with  the NGC~2024S filament being in the process of fragmenting into cores.

\item We modeled the velocity field of the filament and compared the resulting synthetic velocity structure functions with that observed in H$^{13}$CO$^+$. In our toy model, we took the following three velocity components into account: a transverse velocity gradient, a longitudinal velocity gradient, and a longitudinal oscillation caused by fragmentation. The velocity structure function of the Nobeyama H$^{13}$CO$^+$ centroid velocity data shows a longitudinal oscillation pattern reminiscent of that produced by fragmentation in the model. This suggests that our observations are partly tracing core-forming motions resulting from fragmentation of the NGC~2024S filament into cores. The real physical structure of the NGC2024S filament is nevertheless more complex than the prediction of our simple toy model.

\item The average core mass observed in NGC~2024S agrees well with the effective Bonnor-Ebert mass in the filament. Based on a correlation between typical core mass and mass per unit length observed for the Taurus B211/B213, X-shaped nebula in California, NGC~2024S, and  NGC~6334 filaments, we suggest that higher-mass cores may form in higher $M_{\rm line}$ filaments.

\end{itemize}

\begin{acknowledgements}
The 45-m radio telescope is operated by Nobeyama Radio Observatory, a branch of National Astronomical Observatory of Japan. The authors are grateful to  B. Ladjelate for useful discussions. This work was supported by the ANR-11-BS56-010 project ``STARFICH" and the European Research Council under the European Union's Seventh Framework Programme (ERC Advanced Grant Agreement no. 291294 --  `ORISTARS'). YS also received support from the ANR (project NIKA2SKY, grant agreement ANR-15-CE31-0017). This work was supported by NAOJ ALMA Scientific Research Grant Numbers 2017-04A and JSPS KAKENHI Grant Numbers JP19K23463, JP20K04035, and JP21H00057. We also acknowledge support from ``Ile de France'' regional funding (DIM-ACAV$+$ Program) and from the French national programs of CNRS/INSU on stellar and ISM physics (PNPS and PCMI). 
\end{acknowledgements}

%-------------------------------------------------------------------

\bibliographystyle{aa}
\bibliography{Shimajiri}

%\appendix
\begin{appendix}
\section{Complementary figures}
\begin{figure}
\centering
\includegraphics[angle=0,width=8cm]{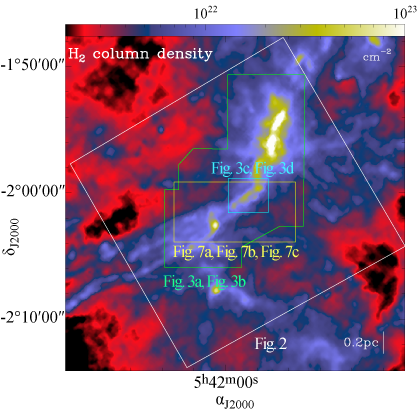}
\caption{$Herschel$ column density map indicating the area of each figure. A white box indicates the area shown in Fig. \ref{fig2}. A green polygon indicates the area shown in Fig. \ref{fig4}a and Fig. \ref{fig4}b. A yellow box indicates the area shown in Fig. \ref{fig7}a, Fig. \ref{fig7}b, and Fig. \ref{fig7}c. A cyan box indicates the area shown in Fig. \ref{fig4}c and Fig. \ref{fig4}d. 
}
\label{figA1}
\end{figure}

Figure \ref{figA1} indicates the area of each figure used in this paper. Figure \ref{figA2} shows a comparison of the velocity channel maps of i) NOEMA H$^{13}$CO$^+$(1--0) data, ii) Nobeyama H$^{13}$CO$^+$(1--0) data, iii) data combined the NOEMA data with the Nobeyama data (hereafter, called NOEMA+45m data), and iv) NOEMA+45m data smoothed to the angular resolution of the Nobeyama H$^{13}$CO$^+$(1--0) data. Figure \ref{figA3} shows a comparison of the {\it Herschel} H$_2$ column density map, Nobeyama 45m C$^{18}$O integrated intensity map, and Nobeyama 45m H$^{13}$CO$^+$ integrated intensity map. Figure~\ref{figA4} shows a pixel-to-pixel correlation plot between Nobeyama H$^{13}$CO$^+$, Nobeyama C$^{18}$O,  and {\it Herschel} H$_2$ column density data.

%Figure A.1

\begin{figure*}
\centering
\includegraphics[angle=0,width=17cm]{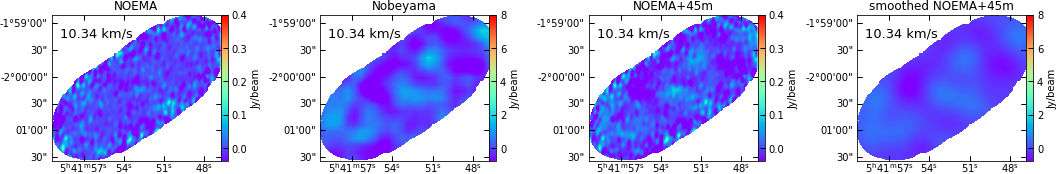}
\includegraphics[angle=0,width=17cm]{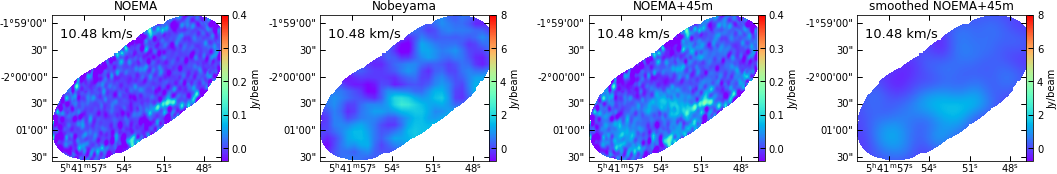}
\includegraphics[angle=0,width=17cm]{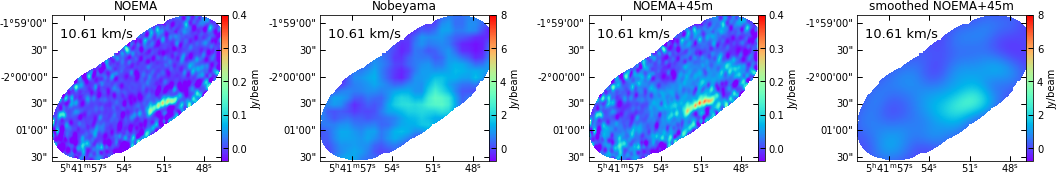}
\includegraphics[angle=0,width=17cm]{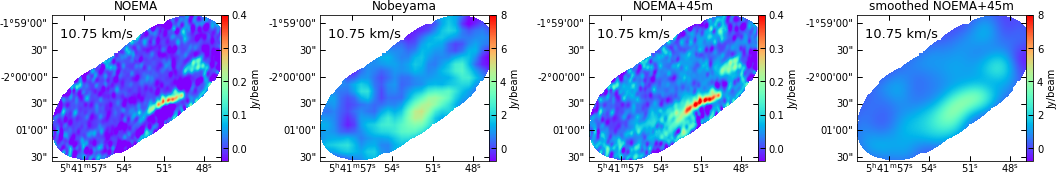}
\includegraphics[angle=0,width=17cm]{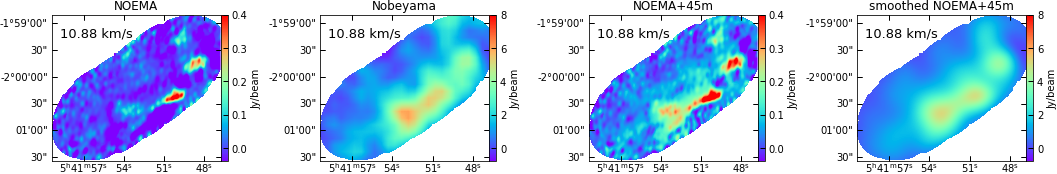}
\includegraphics[angle=0,width=17cm]{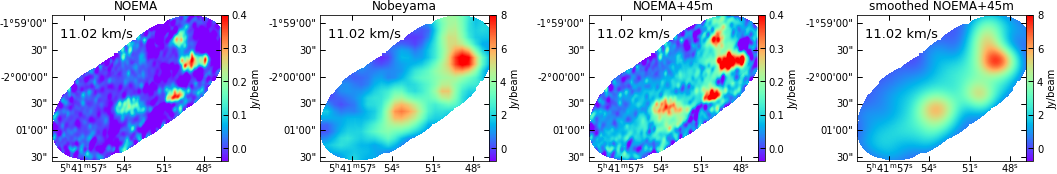}
\includegraphics[angle=0,width=17cm]{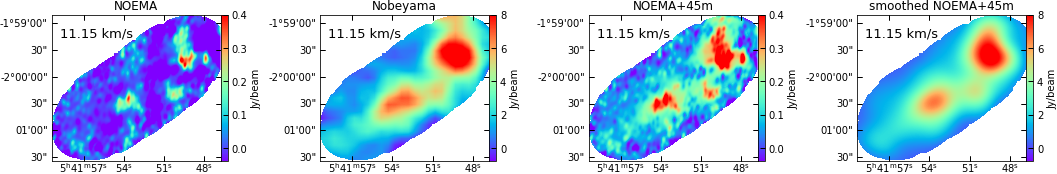}
\caption{Comparison of the velocity channel maps of (left) NOEMA H$^{13}$CO$^+$(1--0) data, (second from the left) Nobeyama H$^{13}$CO$^+$(1--0) data, (third from left) NOEMA+45m data, and (right) smoothed NOEMA+45m data. At the top left of each panel, the velocity is indicated.}
\label{figA2}
\end{figure*}

\addtocounter{figure}{-1}
\begin{figure*}
\centering
\includegraphics[angle=0,width=17cm]{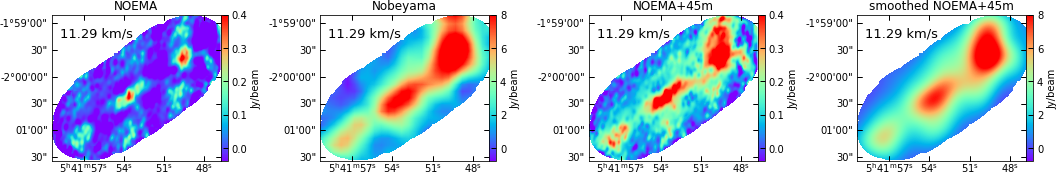}
\includegraphics[angle=0,width=17cm]{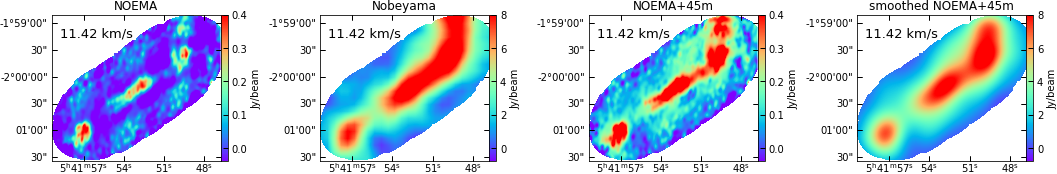}
\includegraphics[angle=0,width=17cm]{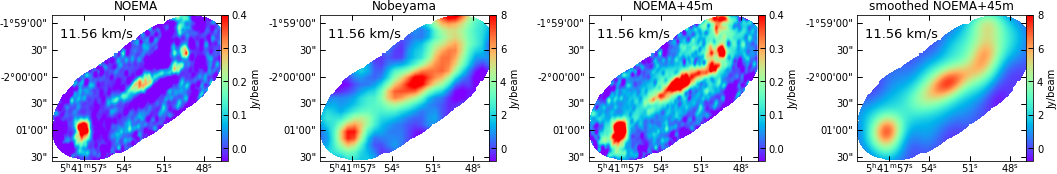}
\includegraphics[angle=0,width=17cm]{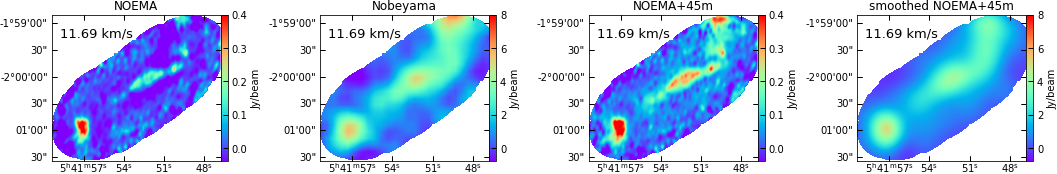}
\includegraphics[angle=0,width=17cm]{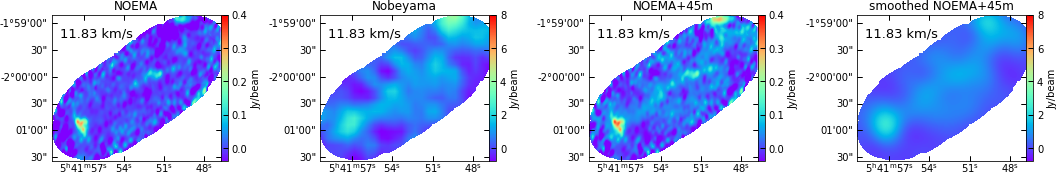}
\includegraphics[angle=0,width=17cm]{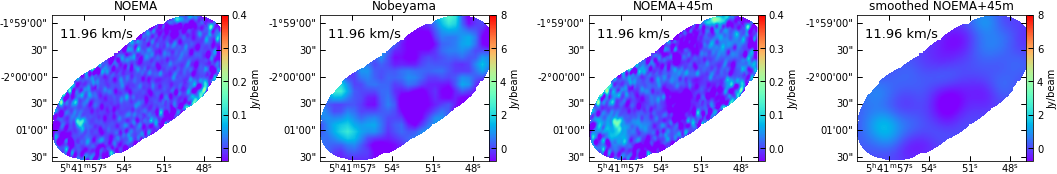}
\includegraphics[angle=0,width=17cm]{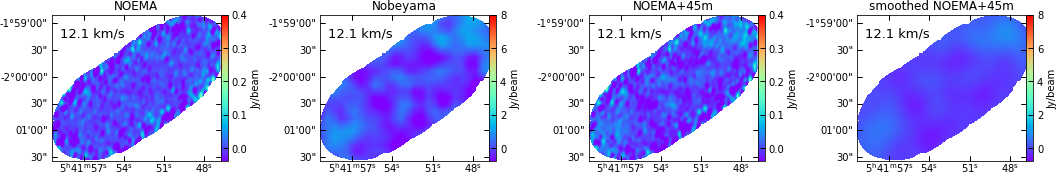}
\caption{(continued)}
\end{figure*}

%Figure 3
\begin{figure*}
\centering
\includegraphics[angle=0,width=16cm]{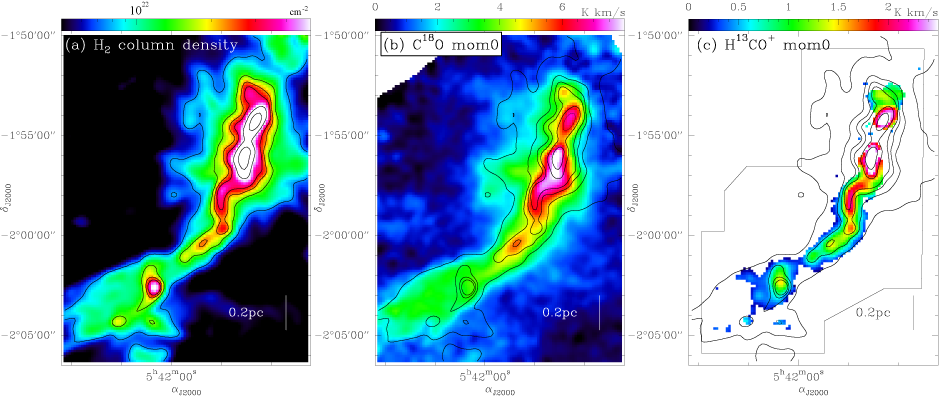}
\caption{Comparison of ($a$) {\it Herschel} H$_2$ column density, ($b$) Nobeyama 45m C$^{18}$O integrated intensity, and ($c$) Nobeyama 45m H$^{13}$CO$^+$ (1--0) integrated intensity maps with an angular resolution of 30$\arcsec$. The panel a is the same as Fig. \ref{fig1}, but the angular resolution is smooth to be 30$\arcsec$. The panel b is the same as Fig. \ref{fig2}m, but the angular resolution is smooth to be 30$\arcsec$. The panel c is the same as Fig. \ref{fig4}b. The black contours in each panel indicate the $A_{\rm V}$ column density levels of 8, 16, 24, 32, 64, 128, and 256 mag (assuming $N_{\rm H_2}$ /$A_{\rm V}$ = 0.94 $\times$ 10$^{21}$ cm$^{-2}$, \cite{Bohlin78}).
}
\label{figA3}
\end{figure*}

\begin{figure*}
\centering
\includegraphics[angle=0,width=8cm]{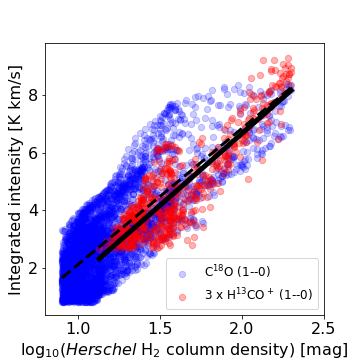}
\caption{Pixel to pixel correlation between H$^{13}$CO$^+$ (1--0) and C$^{18}$O (1--0) integrated intensities in K km s$^{-1}$ and {\it Herschel} H$_2$ column density in mag (assuming $N_{\rm H_2}$ /$A_{\rm V}$ = 0.94 $\times$ 10$^{21}$ cm$^{-2}$, \cite{Bohlin78}). The blue and red points indicate the correlation between C$^{18}$O (1--0) integrated intensity and {\it Herschel} H$_2$ column density and between H$^{13}$CO$^+$ (1--0) integrated intensity and {\it Herschel} H$_2$ column density, respectively. The dashed and solid lines indicate the best-fit result for the C$^{18}$O - {\it Herschel} H$_2$ column density correlation and for the H$^{13}$CO$^+$ - {\it Herschel} H$_2$ column density correlation.}
\label{figA4}
\end{figure*}

\section{Comparison of the median radial profiles between the 8$\arcsec$-resolution ArT$\acute{e}$MiS+$Herschel$ and 18$\arcsec$.2-resolution $Herschel$ column density maps}

For comparison with Fig.~\ref{fig6} obtained from 8$\arcsec$-resolution ArT$\acute{e}$MiS+$Herschel$ data, Fig.~\ref{figB2} 
shows the median radial profiles of the NGC~2024S filament derived from the 18$\arcsec$.2-resolution $Herschel$ column density map of Fig.~\ref{fig1}.
Despite the difference in angular resolution (Table \ref{table_filament_width}), the $Herschel$  and ArT$\acute{e}$MiS+$Herschel$ radial profiles 
and corresponding width measurements are consistent with each other (see \citealp{Andre+2022} for a detailed discussion for the effect of resolution on filament 
width measurements).

\begin{figure}
\centering
\includegraphics[angle=0,width=8cm]{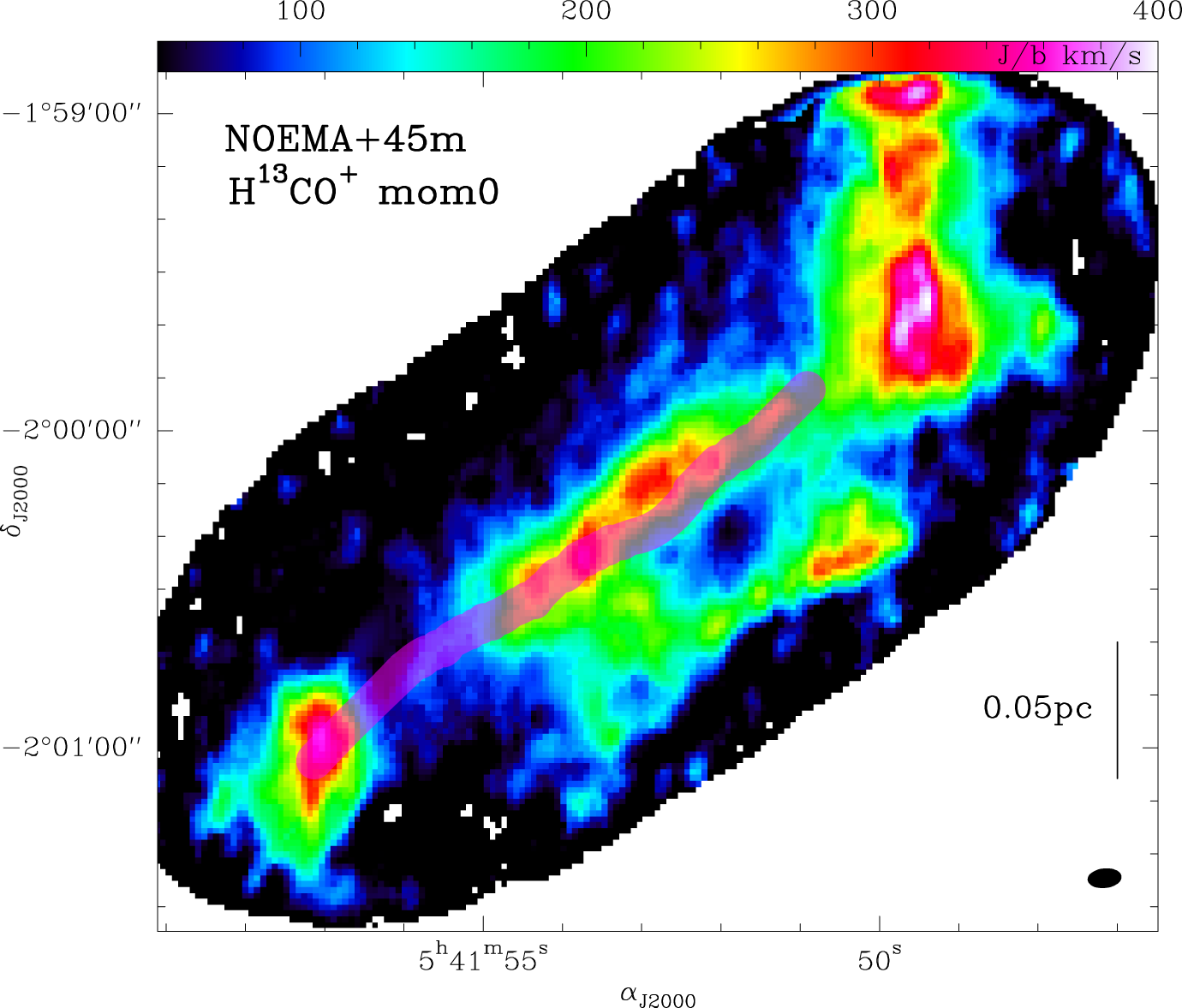}
\caption{Filament crest on the NOEMA+45m H$^{13}$CO$^+$(1--0) integrated intensity maps. The filament crest is determined by DisPerSE algorithm \citep{Sousbie11,Sousbie11b, Arzoumanian11} and are used for producing the median radial profiles shown in Figs. \ref{fig6} and \ref{figB2}. 
}
\label{figB1}
\end{figure}

\begin{figure*}
\centering
\includegraphics[angle=0,width=9cm]{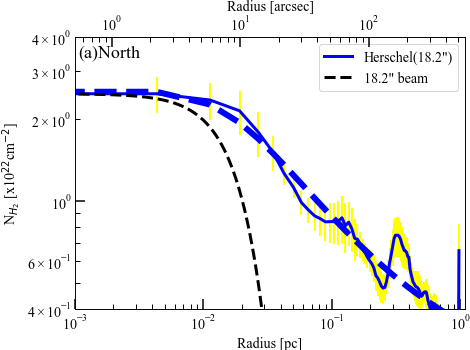}
\includegraphics[angle=0,width=9cm]{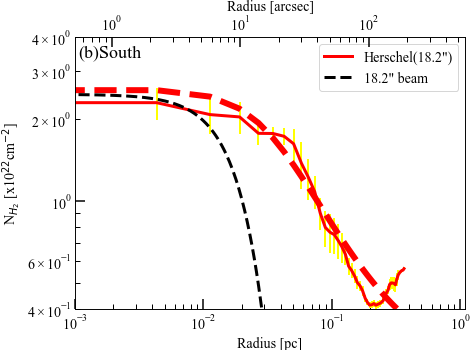}
\caption{Median radial $Herschel$ column density profiles for the (a) northeastern and (b) southwestern side of the NGC~2024S filament. The defined crest of the filament is shown in Fig. \ref{figB1}.
The black dashed curves indicate the angular resolution of the $Herschel$ column density map (18$\arcsec$.2). The dashed curves show the best-fit Plummer mode. The yellow bars show the dispersion ($\pm$1$\sigma$) of the distribution of the radial profile along the filament. The area affected by the secondary component seen in NOEMA H$^{13}$CO$^+$ is avoided to produce the median radial profile for the southwestern side of the NGC~2024S filament (see Sect. \ref{sect:high-reso_map}).
}
\label{figB2}
\end{figure*}

\end{appendix}
\end{document}